\documentclass[letter]{aa}
\usepackage[varg]{txfonts}
\usepackage{graphicx}
\usepackage{natbib,twoopt}
\usepackage[breaklinks=true]{hyperref} 
\usepackage{lscape}
\DeclareUnicodeCharacter{021B}{\textcommabelow t} 
\bibpunct{(}{)}{;}{a}{}{,} 
\makeatletter
 \newcommandtwoopt{\citeads}[3][][]{\href{https://ui.adsabs.harvard.edu/abs/#3/abstract}%
 {\def\hyper@linkstart##1##2{}%
 \let\hyper@linkend\@empty\citealp[#1][#2]{#3}}}
 \newcommandtwoopt{\citepads}[3][][]{\href{https://ui.adsabs.harvard.edu/abs/#3/abstract}%
 {\def\hyper@linkstart##1##2{}%
 \let\hyper@linkend\@empty\citep[#1][#2]{#3}}}
 \newcommandtwoopt{\citetads}[3][][]{\href{https://ui.adsabs.harvard.edu/abs/#3/abstract}%
 {\def\hyper@linkstart##1##2{}%
 \let\hyper@linkend\@empty\citet[#1][#2]{#3}}}
 \newcommandtwoopt{\citeyearads}[3][][]%
 {\href{https://ui.adsabs.harvard.edu/abs/#3/abstract}
 {\def\hyper@linkstart##1##2{}%
 \let\hyper@linkend\@empty\citeyear[#1][#2]{#3}}}
\makeatother

\begin{document}
   \title{Shepherding Miorita and its flock: A group of near-Earth asteroids driven by 
          apsidal and von Zeipel-Lidov-Kozai secular resonances\thanks{Based on 
          observations made with the Isaac Newton Telescope (INT), in the Spanish 
          Observatorio del Roque de los Muchachos of the Instituto de Astrof\'{\i}sica 
          de Canarias (IAC, program ID 136-INT9/14A).}
         }
   \subtitle{A source of low-perihelion asteroids}
   \author{R.~de~la~Fuente Marcos\inst{1}
            \and
           C.~de~la~Fuente Marcos\inst{2}
            \and
           O. V\u{a}duvescu\inst{3}
          }
   \authorrunning{R. de la Fuente Marcos et al.}
   \titlerunning{Miorita et al.: A doubly secularly resonant group} 
   \offprints{R. de la Fuente Marcos, \email{rauldelafuentemarcos@ucm.es}}
   \institute{AEGORA Research Group,
              Facultad de Ciencias Matem\'aticas,
              Universidad Complutense de Madrid,
              Ciudad Universitaria, E-28040 Madrid, Spain
              \and
              Universidad Complutense de Madrid,
              Ciudad Universitaria, E-28040 Madrid, Spain
              \and  
              University of Craiova, 
              Str. A. I. Cuza 13, 200585, Craiova, Romania
             }
   \date{Received 18 June 2025 / Accepted 28 July 2025}
   \abstract
      {Secular resonances can control the dynamical evolution of near-Earth 
       asteroids (NEAs) and, in some cases, lead to increased orbital 
       stability. Asteroid 622577 Miorita (2014~LU$_{14}$) was the first NEA 
       found by the Isaac Newton Telescope (INT) and exhibits unusual 
       dynamical traits although it approaches Venus, Earth, and Mars at 
       relatively close range.
       }
      {Here, we investigate the orbital context of Miorita and search for 
       possible dynamical analogs within the NEA population. 
       }
      {We studied the orbital evolution of Miorita using direct $N$-body 
       calculations. We used the NEOMOD~3 orbital distribution model to verify 
       our conclusions. Observational data were obtained with INT's Wide Field 
       Camera. 
       }
      {Miorita is subjected to a von Zeipel-Lidov-Kozai secular resonance, but 
       it is also in a near apsidal resonance, both controlled by Jupiter. We 
       identified a group of dynamical analogs of Miorita that includes 387668 
       (2002~SZ), 2004~US$_{1}$, 299582 (2006~GQ$_{2}$), and 2018~AC$_{4}$. 
       Miorita-like orbits can evolve into metastable, low-perihelion 
       trajectories driven by apsidal and von Zeipel-Lidov-Kozai secular 
       resonances like those of 504181 (2006~TC) and 482798 (2013~QK$_{48}$). 
       Objects in such paths may end up drawn into the Sun. 
       }
      {Concurrent secular resonances tend to stabilize the orbits of these 
       asteroids as they are protected against collision with Earth and other 
       inner planets by the resonances. This group signals the existence of 
       an active dynamical pathway capable of inserting NEAs in comet-like 
       orbits. NEOMOD~3 gives a low probability for the existence of NEAs like 
       Miorita, 504181 or 482798.
       }


   \maketitle
 
   \nolinenumbers  

   \section{Introduction\label{Intro}}
      Jupiter destabilizes the main asteroid belt, keeping dynamical pathways leading into Earth-crossing orbits open (see, e.g., 
      \citealt{2023AJ....166...55N,2024Icar..41115922N,2024Icar..41716110N,2025Icar..42516316D}). In this way, the near-Earth asteroid (NEA) 
      population gathers new members that replace those lost to the Sun, to the inner planets, or ejected towards the outer regions of the Solar 
      System. However, Jupiter's gravitational perturbation not only makes asteroid orbits unstable, in certain cases it may also turn unstable 
      orbits into stable ones, decreasing the probability of planetary encounters and delaying/preventing eventual collisions 
      \citep{1989Icar...78..212M}. 

      Mean-motion resonances between asteroids and Jupiter produce the Kirkwood gaps in the main asteroid belt (see, e.g., 
      \citealt{1983Natur.301..201D}). In addition, secular resonances induced by Jupiter can sometimes make orbits unstable but, in other cases, these 
      secular perturbations may lead to enhanced orbital stability (see, e.g., \citealt{1995Icar..117...45F,2017MNRAS.468.4719V,2024CeMDA.136...17V}). 
      Increased local orbital stability causes the formation of dynamical groups.
       
      Asteroid 622577 Miorita (2014~LU$_{14}$) was the first NEA found from La Palma by the Isaac Newton Telescope (INT, \citealt{2014MPEC....L...33T,
      2015MNRAS.449.1614V}); its assigned name is an old Romanian pastoral ballad.\footnote{``Miori{\textcommabelow t}a is a unique Romanian pastoral 
      myth-ballad. A little ewe with supernatural powers denounces the evil plan of two shepherds to murder her master. But this shepherd resigns 
      himself to his fate, asking to be buried by the sheepfold and imagining his cosmic wedding witnessed by nature, Sun, Moon, torch stars and one 
      falling star, meaning his death.'' Ref: WGSBN Bull.\ 4, \#11, 22} It exhibits unusual dynamical traits perhaps compatible with orbital 
      stabilization via secular resonances. Here, we use $N$-body simulations to explore its dynamical context and implications. This Letter is 
      organized as follows. In Sect.~\ref{Data}, we introduce the background of our work, and present the data and tools used in our analyses. In 
      Sect.~\ref{Results}, we explore the orbital properties and resonant context of Miorita, and its probable origin together with those of a sample 
      of dynamical analogs. In Sect.~\ref{Discussion}, we discuss our results and Sect.~\ref{Conclusions} summarizes our conclusions. Appendices 
      include supporting material.

   \section{Context, methods, and data\label{Data}}
      In this section, we revisit dynamical concepts that are later used in our analysis. Software tools and data are also discussed here. 

      \subsection{Dynamics background}
         Secular resonances are organized into two different types, those affecting the precession of, e.g., the line of apsides and a resonant 
         mechanism that exchanges eccentricity, $e$, and inclination, $i$, driving antiparallel oscillations or librations of these two orbital 
         elements, the von Zeipel-Lidov-Kozai secular resonance \citep{1910AN....183..345V,1962AJ.....67..591K,1962P&SS....9..719L, 
         2016ARA&A..54..441N,2018CeMDA.130....4S,2019MEEP....7....1I,2023MNRAS.522..937T}. An apsidal resonance between an asteroid and a planet 
         appears when there is a periodic libration of the relative apsidal longitude or relative longitude of perihelion, $\Delta \varpi = \varpi - 
         \varpi_{\rm P}$ (see, e.g., \citealt{1985CeMec..36..391N,1986A&A...170..138S}), where $\varpi = \Omega + \omega$ is the longitude of 
         perihelion of the asteroid, $\varpi_{\rm P} = \Omega_{\rm P} + \omega_{\rm P}$ is the one of the planet, $\Omega$ is the longitude of the 
         ascending node of the asteroid and $\omega$ is its argument of perihelion, $\Omega_{\rm P}$ and $\omega_{\rm P}$ are the equivalent 
         parameters for the planet (see, e.g., \citealt{1999ssd..book.....M}). An extensive map of secular resonances in the NEA region was computed 
         by \citet{2023A&A...672A..39F}. Secular resonances play a prominent role on the dynamical evolution of extrasolar planetary systems (see, 
         e.g., \citealt{2003ApJ...598.1290Z,2009A&A...493..677L,2022A&A...665A..62L}). 

      \subsection{Data, data sources, and tools}
         Apollo-class NEA 622577 Miorita (2014~LU$_{14}$) was initially reported on June 2, 2014, by the EUROpean Near Earth Asteroids Research 
         (EURONEAR, \citealt{2008P&SS...56.1913V})\footnote{\href{http://www.euronear.org/}{http://www.euronear.org/}} project observing with the 
         Isaac Newton Telescope (INT) from La Palma; it was announced on June 6 with the provisional designation 2014~LU$_{14}$ 
         \citep{2014MPEC....L...33T}. The initial orbit solution was further improved using observations obtained by EURONEAR (see discussion in 
         Appendix~\ref{INTdata}) and other collaborations. Its orbital solution in Table~\ref{elements} is currently based on 95 observations with an 
         observational timespan of 3644~d, and it is referred to epoch JD 2460800.5 TDB. It was retrieved from Jet Propulsion Laboratory's (JPL) 
         Small-Body Database (SBDB)\footnote{\href{https://ssd.jpl.nasa.gov/tools/sbdb\_lookup.html\#/}
         {https://ssd.jpl.nasa.gov/tools/sbdb\_lookup.html\#/}} provided by the Solar System Dynamics Group (SSDG, 
         \citealt{2011jsrs.conf...87G,2015IAUGA..2256293G}).\footnote{\href{https://ssd.jpl.nasa.gov/}{https://ssd.jpl.nasa.gov/}} For Miorita, the 
         Spitzer mission gives an absolute magnitude of 19.89$\pm$0.33~mag, a diameter of 0.357$^{+0.142}_{-0.074}$~km, and an albedo of 
         0.151$^{+0.100}_{-0.075}$ \citep{2019AJ....158...67G}. The object attracted our attention because it experiences relatively close encounters 
         with Venus, the Earth--Moon system, and Mars.

         The orbit of Miorita crosses those of Venus, the Earth--Moon system, and Mars. Periodic planetary close encounters might make the 
         reconstruction of Miorita's past orbital evolution and the prediction of its future behavior beyond a few kyr difficult. In such cases, the 
         orbital evolution has to be investigated statistically considering the uncertainties of the orbital solution. The calculations needed to 
         study the evolution of Miorita were carried out using a direct $N$-body code described by \citet{2003gnbs.book.....A} and publicly available 
         from the web site of the Institute of Astronomy of the University of 
         Cambridge.\footnote{\href{https://people.ast.cam.ac.uk/~sverre/web/pages/nbody.htm}
         {https://people.ast.cam.ac.uk/~sverre/web/pages/nbody.htm}} This software implements the Hermite numerical integration scheme introduced by 
         \citet{1991ApJ...369..200M}. Details and results from this code can be found in \citet{2012MNRAS.427..728D}. Our 
         physical model included the perturbations by the eight major planets, the Moon, the barycenter of the Pluto-Charon system, and the 
         19 largest asteroids, Ceres, Pallas, Vesta, Hygiea, Euphrosyne, Interamnia, Davida, Herculina, Eunomia, Juno, Psyche, Europa, 
         Thisbe, Iris, Egeria, Diotima, Amphitrite, Sylvia, and Doris. For accurate initial positions and velocities (see 
         Appendix~\ref{Adata}), we used data based on the DE440/441 planetary ephemeris \citep{2021AJ....161..105P} from JPL's SSDG 
         {\tt Horizons} on-line Solar system data and ephemeris computation service.\footnote{\href{https://ssd.jpl.nasa.gov/horizons/}
         {https://ssd.jpl.nasa.gov/horizons/}} Most input data were retrieved from SBDB and {\tt Horizons} using tools provided by the {\tt Python} 
         package {\tt Astroquery} \citep{2019AJ....157...98G} and its {\tt HorizonsClass} 
         class.\footnote{\href{https://astroquery.readthedocs.io/en/latest/jplhorizons/jplhorizons.html}
         {https://astroquery.readthedocs.io/en/latest/jplhorizons/jplhorizons.html}} Figures were produced using {\tt Matplotlib} 
         \citep{2007CSE.....9...90H}. The plausibility of our conclusions was examined within the context of the NEOMOD~3 orbital distribution model
         \citep{2024Icar..41716110N}. NEOMOD~3 is the latest update of the NEOMOD project \citep{2023AJ....166...55N,
         2024Icar..41115922N}. 

   \section{Results: Orbital evolution\label{Results}}
\begin{figure*}
        \centering
         \includegraphics[width=0.32\linewidth]{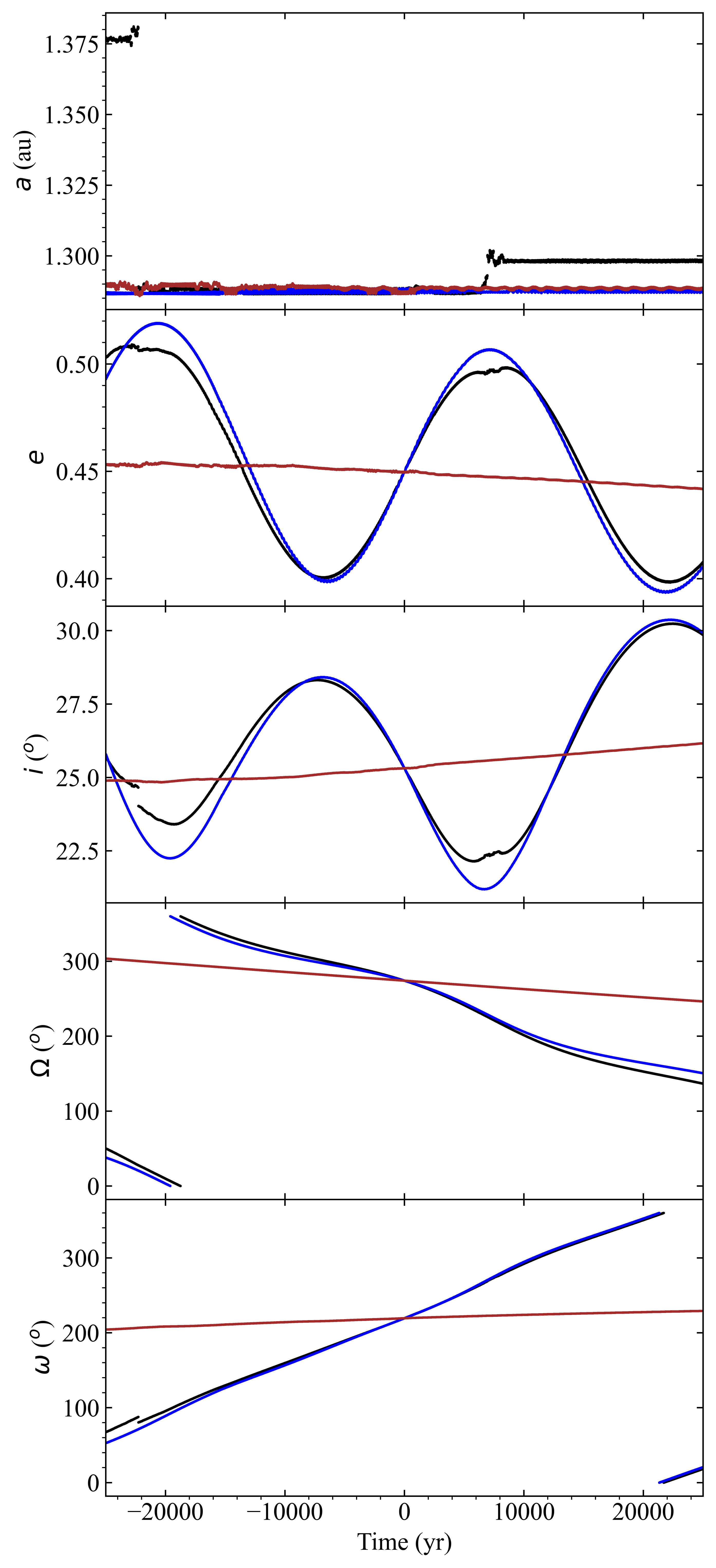}
         \includegraphics[width=0.32\linewidth]{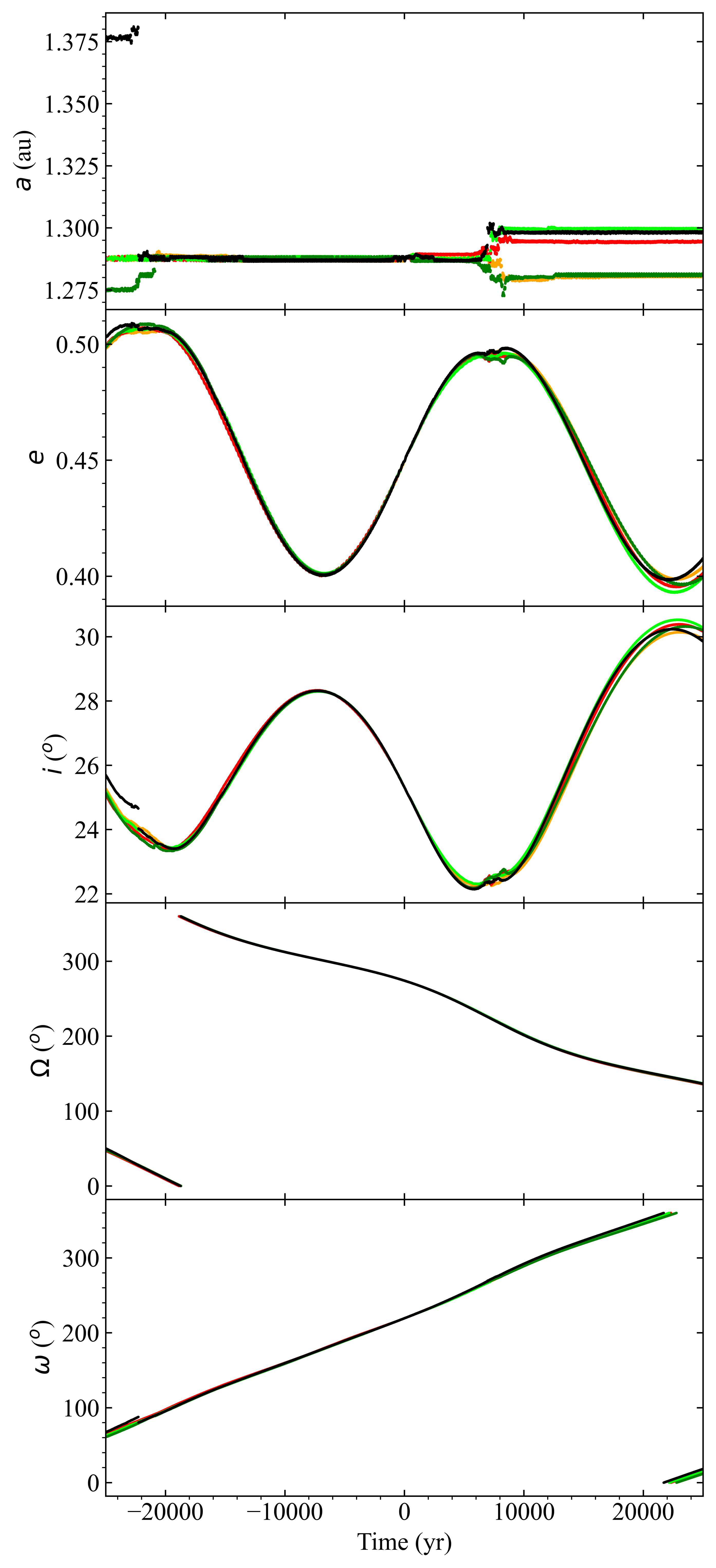}
         \includegraphics[width=0.315\linewidth]{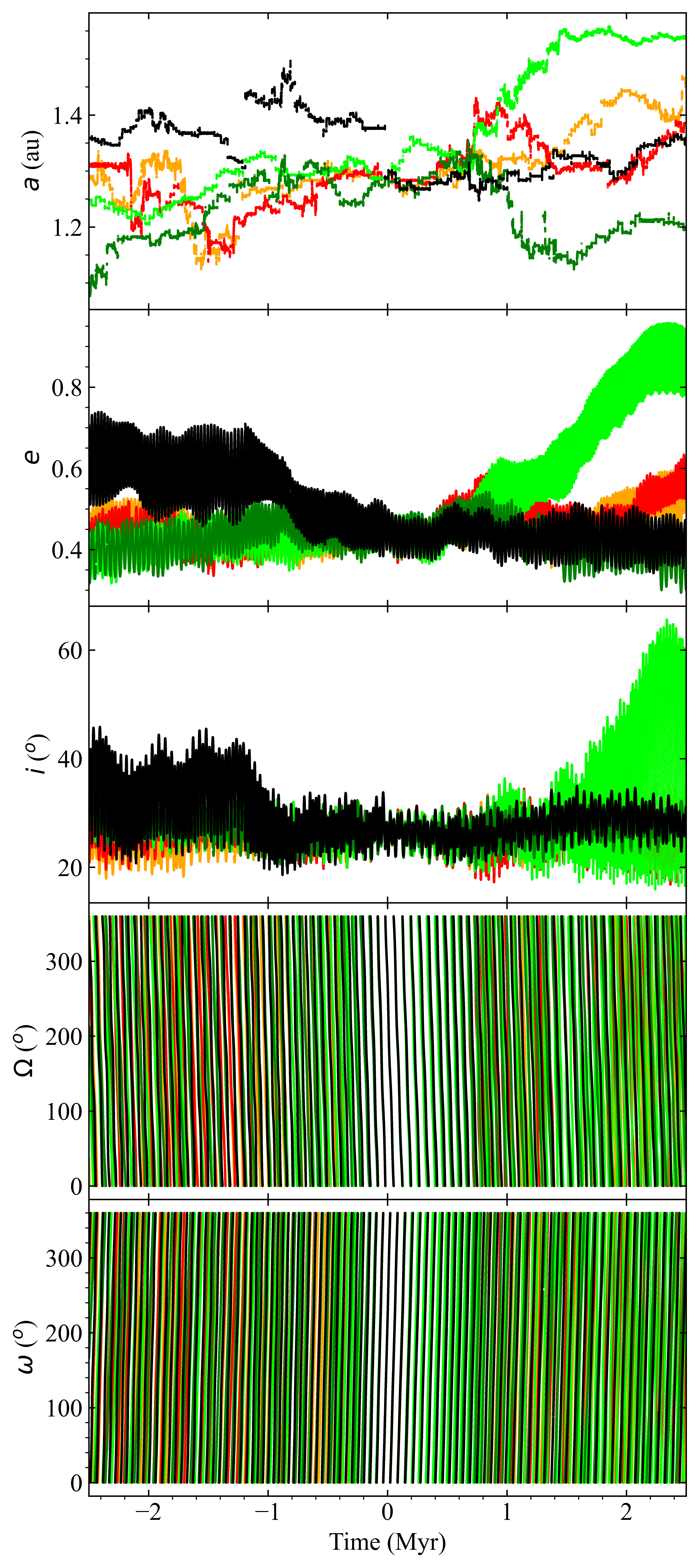}
         \caption{Orbital evolution of 622577 Miorita (2014~LU$_{14}$). {\it Left panels:} Results of a simulation of the nominal orbit and the full 
                  physical model (in black), after removing the Earth--Moon system (in blue), and after removing Jupiter (in brown). {\it Central 
                  panels:} Short-term results of the full physical model and the nominal orbit (in black), those of control orbits with state vectors 
                  separated $\pm3\sigma$ from the nominal ones in lime/green, and $\pm9\sigma$ in orange/red. {\it Right panels:} Integrations in the 
                  central panels but on a longer timespan. Time evolution of the value of the semimajor axis, $a$ (top panels), eccentricity, $e$ 
                  (second to top), inclination, $i$ (middle), the longitude of the ascending node, $\Omega$ (second to bottom), and the argument of 
                  perihelion, $\omega$ (bottom). The origin of time is the epoch 2460800.5~JD Barycentric Dynamical Time (2025-May-05.0 00:00:00.0 
                  TDB) and the output cadence is 36.525~d for the left and central panels, and 25~yr for the right panels. The source of the input 
                  data is JPL's {\tt Horizons}.
                 }
         \label{Miorita}
      \end{figure*}

      Here, we use $N$-body simulations to assess the current dynamical status of 622577 Miorita (2014~LU$_{14}$) and its implications. 
      Figure~\ref{Miorita} shows the time evolution of the orbital elements of Miorita and summarizes the results of our calculations. The central
      panels focus on the short-term evolution of the nominal orbit and those of representative control orbits or clones with state vectors 
      (Cartesian coordinates and velocities, see Table~\ref{vectorMiorita} in Appendix~\ref{Adata}) well away from the nominal one, up to 
      $\pm9\sigma$ from the nominal orbital solution in Table~\ref{elements}. Given the reasonable good quality of this orbital solution, even a 
      deviation of $\pm9\sigma$ in the values of the orbital elements gives a clone with initial conditions separated $<$0.001\% from the nominal 
      ones. The central panels of Fig.~\ref{Miorita} show that control orbits that start arbitrarily close tend to diverge on a timescale of a few 
      kyr, due to the effect of close encounters with Earth. In addition, the panels of $e$ and $i$ display anticorrelated librations of these two 
      orbital elements with a period of 30,000~yr, signaling the presence of the von Zeipel-Lidov-Kozai secular resonance 
      \citep{1910AN....183..345V,1962AJ.....67..591K,1962P&SS....9..719L}. All the control orbits studied (10$^{3}$) exhibit anticorrelated 
      periodic libration of eccentricity and inclination.

      To investigate the sources of orbital divergence (chaotic evolution) and von Zeipel-Lidov-Kozai oscillations, the nominal orbit was 
      integrated using physical models that removed the Earth--Moon system or Jupiter; our results are shown on the left panels of 
      Fig.~\ref{Miorita}. Looking at the evolution of the nominal orbit and the full physical model (in black), and that of the model with the 
      Earth--Moon system removed (in blue), we observe that the kinks in the evolution of $e$ and $i$ dissapear, a confirmation that close 
      planetary encounters are responsible for the onset of the observed chaotic behavior. In contrast, the removal of Jupiter (in brown) has a 
      dramatic effect on the dynamical evolution of Miorita, the von Zeipel-Lidov-Kozai oscillations dissapear. In general, Jupiter fully controls 
      the evolution of the shape ($e$) and the orientation in space ($i$, $\Omega$, and $\omega$) of the orbit of Miorita.  

\begin{figure}
        \centering
         \includegraphics[width=\linewidth]{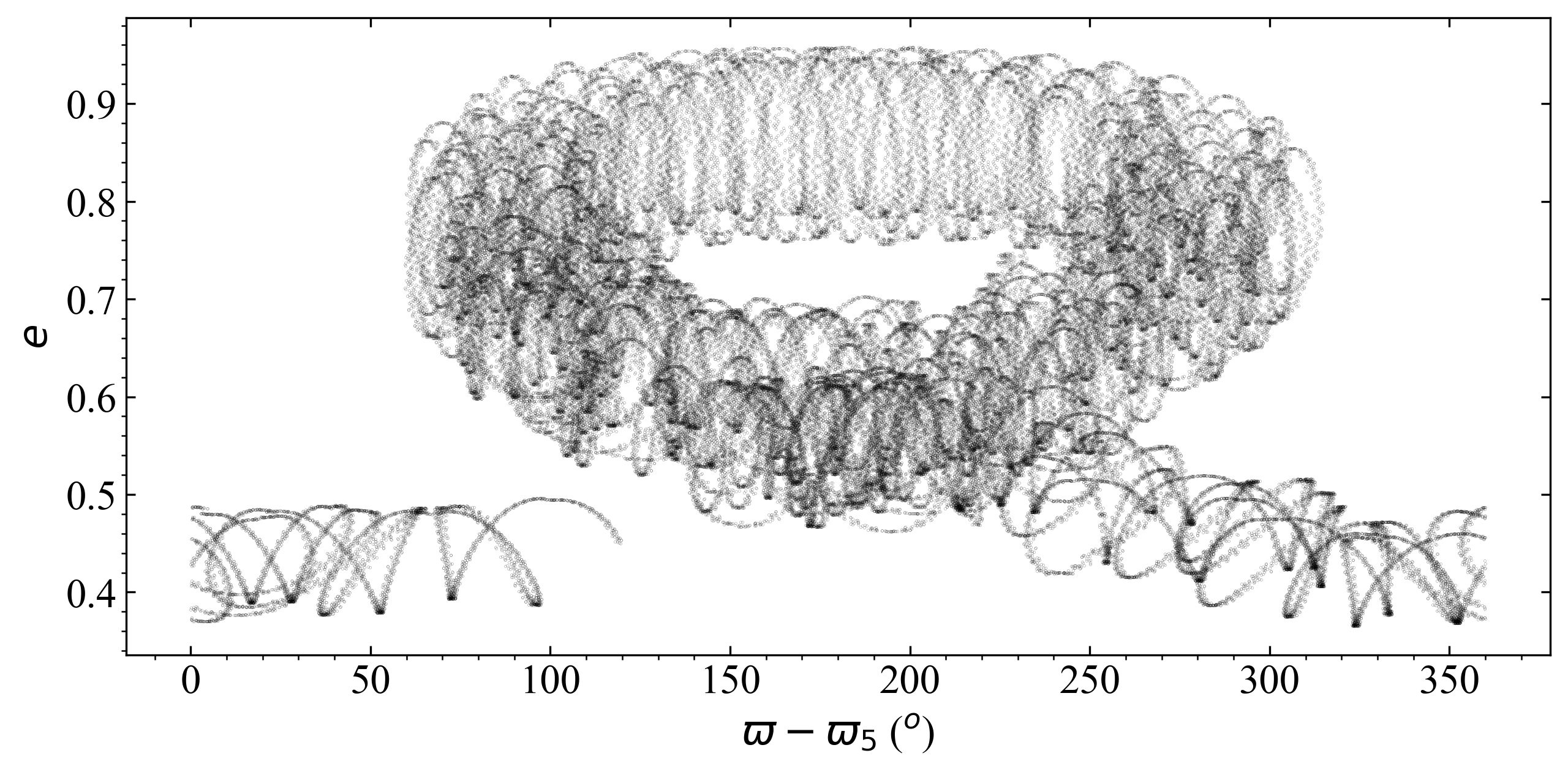}
         \includegraphics[width=\linewidth]{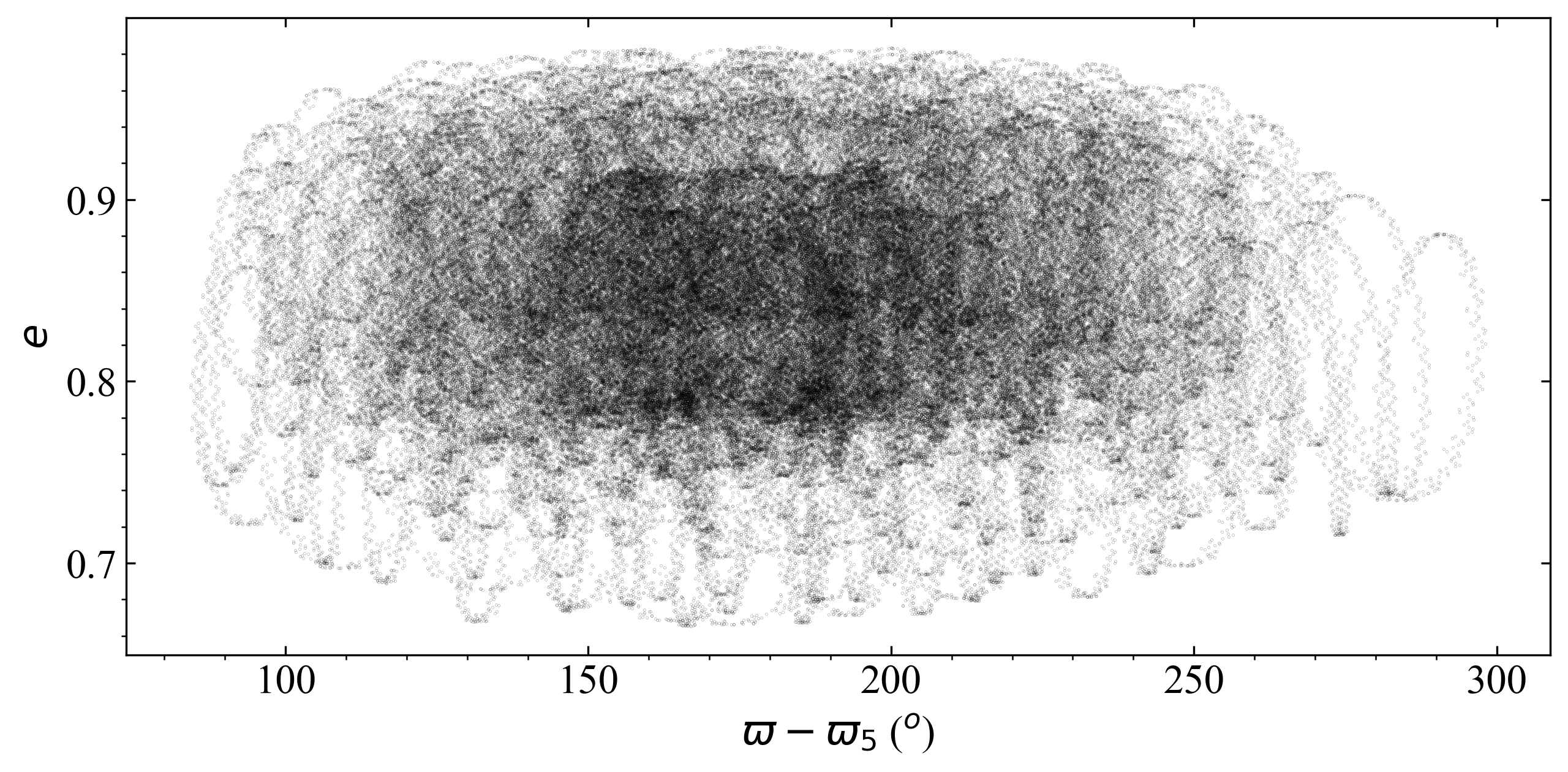}
         \includegraphics[width=\linewidth]{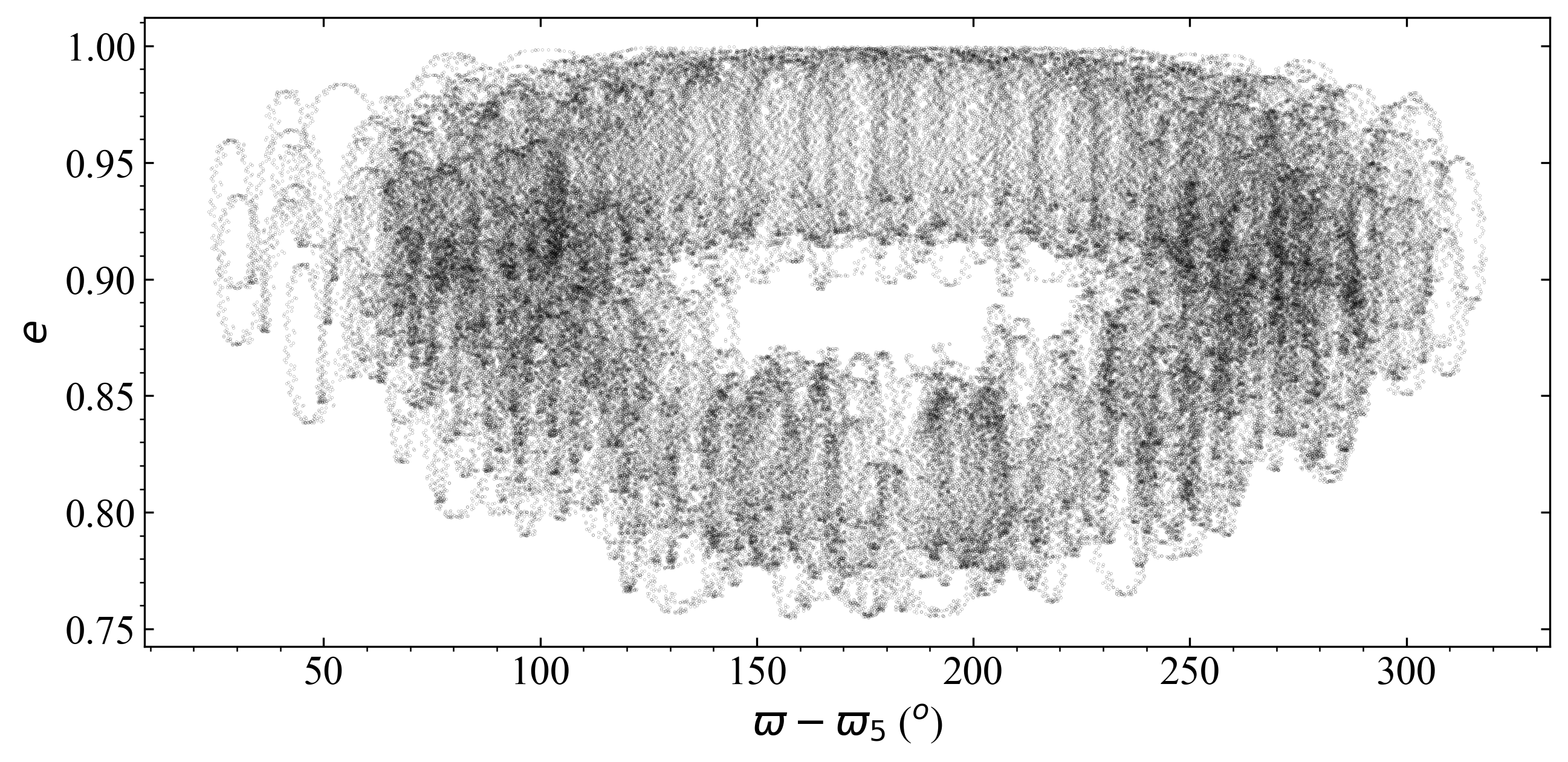}
         \caption{Variation of the eccentricity with the relative (to Jupiter) apsidal longitude $\varpi - \varpi_{\rm 5}$. {\it Top panel:} Results
                  for the control orbit of Miorita with state vectors separated $-3\sigma$ from the nominal ones. {\it Middle panel:} Results for 
                  504181 (2006~TC). {\it Bottom panel:} Results for 482798 (2013~QK$_{48}$). The source of the input data is JPL's {\tt Horizons}.
                 }
         \label{Miorita-apsidal}
      \end{figure}

      Figure~\ref{Miorita}, right panels, shows the results of longer integrations back into the past and forward into the future. These 
      integrations delve deeper into the evolution displayed in the central panels. Although Miorita's orbital evolution is conspicuously chaotic,
      it is also clear that it remains confined to a relatively narrow volume of the orbital parameter space. One of the relevant control orbits,
      the one with state vectors separated $-3\sigma$ from the nominal ones (in lime) deserves detailed study as it reaches a region in which 
      additional secular resonances emerge. The long-term evolution of this control orbit is displayed in Figs.~\ref{apsidal} and 
      \ref{Miorita-apsidal}. In addition to the short-period anticorrelated oscillations of $e$ and $i$, a long-period ($\sim$2~Myr) oscillation 
      emerges that is triggered by the onset of an apsidal resonance with Jupiter as shown in Fig.~\ref{Miorita-apsidal}, top panel. The resonance 
      is about $\varpi - \varpi_{\rm 5} = 180{\degr}$ so the perihelia of Jupiter and Miorita are antialigned, when Miorita reaches perihelion, 
      Jupiter is at aphelion and vice versa. From Figs.~\ref{Miorita-apsidalnu5} and \ref{Miorita-apsidalnu5n} of Appendix~\ref{flock}, Miorita
      evolves close or inside this secular resonance. 
 
   \section{Discussion\label{Discussion}}
      An important open question is how frequent Miorita-like orbits are. To estimate the theoretical likelihood of NEAs like 622577 Miorita 
      (2014~LU$_{14}$), we considered the NEOMOD~3 orbital distribution model \citep{2024Icar..41716110N}. This tool can be used to generate synthetic 
      NEA populations suitable to estimate probabilites or test hypotheses, and reject or accept them; the model can only be strictly applied to 
      objects with $H<28$~mag and does not include fragmentations. We have generated a population of 2.75$\times$10$^{8}$ synthetic NEOs of sizes 
      above 0.001~km with the NEOMOD~3 simulator.\footnote{\href{https://www.boulder.swri.edu/~davidn/NEOMOD\_Simulator/}
      {https://www.boulder.swri.edu/$\sim$davidn/NEOMOD\_Simulator/}} For our purposes, NEOMOD~3 generates a set of synthetic NEO instances 
      characterized by $a$, $e$, $i$, and $H$. From this set, the probability (in the usual sense of number of occurrences in a sample of a certain 
      size) of finding one object with values of its orbital elements ($a$, $e$, $i$) matching those of Miorita in Fig.~\ref{Miorita}, central panels 
      --- namely (1.275, 1.375)~au, (0.39, 0.51), and (22, 30){\degr}, respectively --- is 0.0016. 

      Against this theoretical estimate, we can compare the one from real data. Querying JPL's SBDB within the volume of the orbital parameter space 
      defined by the intervals above, we found 34 objects out of 38631 known NEAs, or a probability close to 0.0009, that is somewhat consistent with 
      the one derived from NEOMOD~3 above, and indicative that such orbits are not long-term stable. Examples of dynamical analogs of Miorita are 
      387668 (2002~SZ), 2004~US$_{1}$, 299582 (2006~GQ$_{2}$), and 2018~AC$_{4}$ (see Fig.~\ref{Miorita-flock} in Appendix~\ref{flock}). But 
      Miorita-like orbits may evolve into paths affected by an apsidal resonance with Jupiter. 

      Among NEAs with reliable orbits, we found two in paths similar to the one described in the previous section and affected by an apsidal resonance 
      with Jupiter: 504181 (2006~TC) and 482798 (2013~QK$_{48}$), see Fig.~\ref{Miorita-apsidal}. NEA 504181 evolves within the ranges of $a$, $e$ and 
      $i$, (1.48, 1.67)~au, (0.60, 0.98), and (18, 70){\degr}; for 482798, the equivalent intervals are (1.54, 1.70)~au, (0.75, 1.0), and 
      (10, 90){\degr} (see Appendix~\ref{beyond}). Using NEOMOD~3, the probability of finding one NEA like 504181 is 0.0022, and for 482798 is 
      0.00057. Querying JPL's SBDB, we found for the case of 504181 100$\pm$10 (Poisson counting uncertainty) objects there, out of 38631 known NEAs, 
      or a probability close to 0.0026; for 482798, the associated probability is 0.0009 (33$\pm$6 objects). The region inhabited by 504181 is nearly 
      four times more stable than that of 482798. In fact, 482798-like orbits tend to be lost to the Sun (see Appendix~\ref{beyond}). The existence of 
      such low-perihelion asteroids was predicted by \citet{2023A&A...672A..39F}. 

   \section{Summary and conclusions\label{Conclusions}}
      In this Letter, we presented an investigation of the current dynamical status of Apollo-class NEA 622577 Miorita (2014~LU$_{14}$).   
      Our conclusions can be summarized as follows.
      \begin{enumerate}
         \item We show that Miorita is concurrently engaged in a von Zeipel-Lidov-Kozai secular resonance and a near apsidal resonance, both driven by 
               Jupiter. Concurrent secular resonances tend to stabilize the orbits of these asteroids. 
         \item We find that a number of NEAs move in Miorita-like orbits.
         \item We find that objects in such orbits may become low-perihelion asteroids concurrently engaged into von Zeipel-Lidov-Kozai 
               oscillations and an apsidal resonance, driven by Jupiter. Some known asteroids exhibit this behavior.
      \end{enumerate}
      Our numerical study has confirmed the existence of an active dynamical pathway capable of inserting NEAs in comet-like orbits; this pathway was 
      originally predicted by \citet{2023A&A...672A..39F}. Such objects may pose as extinct-comet impostors.

   \begin{acknowledgements}
      We thank the anonymous referee for a constructive and timely report. RdlFM acknowledges funding from the ``ENIMUS'' Advanced Grant from the 
      European Research Council (ERC) under the European Union's Horizon 2020 research and innovation programme (grant agreement ID 101097905). This 
      work was partially supported by the Spanish `Agencia Estatal de Investigaci\'on (Ministerio de Ciencia e Innovaci\'on)' under grant 
      PID2020-116726RB-I00 /AEI/10.13039/501100011033. Based on observations made with the Isaac Newton Telescope (INT), installed at the Spanish 
      Observatorio del Roque de los Muchachos of the Instituto de Astrof\'{\i}sica de Canarias, on the island of La Palma. In preparation of this 
      Letter, we made use of the NASA Astrophysics Data System, the ASTRO-PH e-print server, and the Minor Planet Center (MPC) data server. 
   \end{acknowledgements}

   \bibliographystyle{aa}

   \begin{appendix}

      \section{Discovery and follow-up observations\label{INTdata}}
         Apollo-class NEA 622577 Miorita (2014~LU$_{14}$) was discovered serendipitously in observations of the proposal C136 (PI: O. Vaduvescu) aimed 
         at recovering poorly studied NEAs by triggering max 1~h/night shots (sometimes including twilight time) using the 2.5~m-INT. Our program 
         lasted five semesters, thus after two and a half years our collaboration recovered and made it possible to improve the orbital solutions of 
         280 faint NEAs (centered around $V\sim22.8$~mag). In addition, we reported positions of about 3,500 known minor planets and another 1,500 
         unknown objects, producing the first nine EURONEAR serendipitous NEA discoveries \citep{2018A&A...609A.105V}.

         As part of our C136 NEA recovery program and during the morning of June 1--2, 2014, we attempted the recovery of three known NEA, including 
         the known target NEA 486001 (2012~MR$_{7}$), which was observed in the morning due to its relatively low Solar elongation 
         ($\epsilon$=117{\degr}). The discovery of Miorita was mainly due to chance, because it happened during the second recovery attempt of the 
         known target, which was repeated due to a pointing error made one night prior.

         Thanks to the possibility of triggering short INT service time slots during most nights when the Wide Field Camera (WFC) was available, we 
         succesfully recovered Miorita on the second night and we could perform follow-up observations during five nights. For 18 days, no other major 
         survey or individual observer was able to report it (MPS~518060, 518856) until it was confirmed from Cerro Tololo and Mount John (after the 
         full moon), then by Pan-STARRS and Mauna Kea, which extended the data-arc to almost three months. Its orbital solution as of June 15, 2025, 
         is shown in Table~\ref{elements}.
      \begin{table}
         \centering
         \fontsize{8}{11pt}\selectfont
         \tabcolsep 0.15truecm
         \caption{\label{elements}Heliocentric Keplerian orbital elements of asteroid 622577 Miorita (2014~LU$_{14}$). 
                 }
         \begin{tabular}{lcc}
            \hline\hline
             Parameter                                         &   &    Miorita                   \\
            \hline
             Semimajor axis, $a$ (au)                          & = &    1.28685458$\pm$0.00000002 \\
             Eccentricity, $e$                                 & = &    0.4496381$\pm$0.0000004   \\
             Inclination, $i$ (\degr)                          & = &   25.31145$\pm$0.00004       \\
             Longitude of the ascending node, $\Omega$ (\degr) & = &  274.08069$\pm$0.00003       \\
             Argument of perihelion, $\omega$ (\degr)          & = &  219.48826$\pm$0.00005       \\
             Mean anomaly, $M$ (\degr)                         & = &  282.88704$\pm$0.00004       \\
             Perihelion, $q$ (au)                              & = &    0.7082358$\pm$0.0000004   \\
             Aphelion, $Q$ (au)                                & = &    1.86547339$\pm$0.00000003 \\
             MOID with the Earth (au)                          & = &    0.197461                  \\
             Absolute magnitude, $H$ (mag)                     & = &   19.71                      \\
             Data-arc span (d)                                 & = & 3644                         \\
             Number of observations                            & = &   95                         \\
            \hline
         \end{tabular}
         \tablefoot{Values include the 1$\sigma$ uncertainty. The orbit of Miorita (solution date, May 26, 2024, 05:47:28 PDT) is referred to epoch JD
                    2460800.5, which corresponds to 0:00 on 2025 May 5 TDB (Barycentric Dynamical Time, J2000.0 ecliptic and equinox). Source: JPL's 
                    SBDB.
                   }
      \end{table}

         \subsection{Image reduction}
            During the entire recovery program, we promptly reduced the images (usually during the following morning) using the GUI version of the 
            {\tt THELI} software\footnote{\href{https://astro.uni-bonn.de/theli/}{https://astro.uni-bonn.de/theli/}} \citep{2007AN....328..701S,2013ApJS..209...21S} which corrects for bias, 
            flat, and field distortion, an issue that is very important for older telescopes and prime focus instruments such as the WFC. The actual 
            observer and {\it THELI} reducer of the three fields observed during the discovery night was Vlad Tudor, an Isaac Newton Group of 
            Telescopes (ING) student and INT support astronomer who promptly reduced all the fields.

            For the actual moving object search and recovery of the NEA targets, a team of a few amateur astronomers used 
            {\tt Astrometrica}\footnote{\href{http://www.astrometrica.at/}{http://www.astrometrica.at/}} \citep{2012ascl.soft03012R} to carefully scan all reduced images, visually 
            blinking and measuring all moving sources, including the main target, plus any other known and unknown asteroids. The actual reducer of 
            the three observed fields where Miorita was first found was Lucian Hudin, an amateur astronomer from Cluj-Napoca, Romania.

            Finally, the NEA targets of the actual main program, plus the known and unknown detections in the observed fields were checked and 
            promptly reported to the Minor Planet Center (MPC), usually during the next day, allowing the PI to eventually plan for the rapid recovery 
            of any new NEA candidates, using the same INT program or asking other members of the EURONEAR 
            network\footnote{\href{http://www.euronear.org/network.php}{http://www.euronear.org/network.php}} with access to other telescopes.

         \subsection{NEA candidate EUHT171 = 622577 Miorita (2014~LU$_{14}$)}
            Lucian Hudin identified 15 moving sources in the targeted field, including the main target NEA 486001 (recovered about 40" away from the 
            official MPC ephemerides), other five known main belt asteroids (MBAs), and nine other unknown detections, which included Miorita. In 
            order to perform quality control checks, during the entire program, we used the MPC {\tt NEO Rating} 
            tool\footnote{\href{https://minorplanetcenter.net/iau/NEO/PossNEO.html}{https://minorplanetcenter.net/iau/NEO/PossNEO.html}} to check for possible NEO candidacy of all unknown detections. Miorita 
            had the highest rating out of the 15 detections in this field (99\%), then 35123 (74\%, actually an MBA), then EUHT173 (71\%).

\begin{figure}
        \centering
         \includegraphics[width=\linewidth]{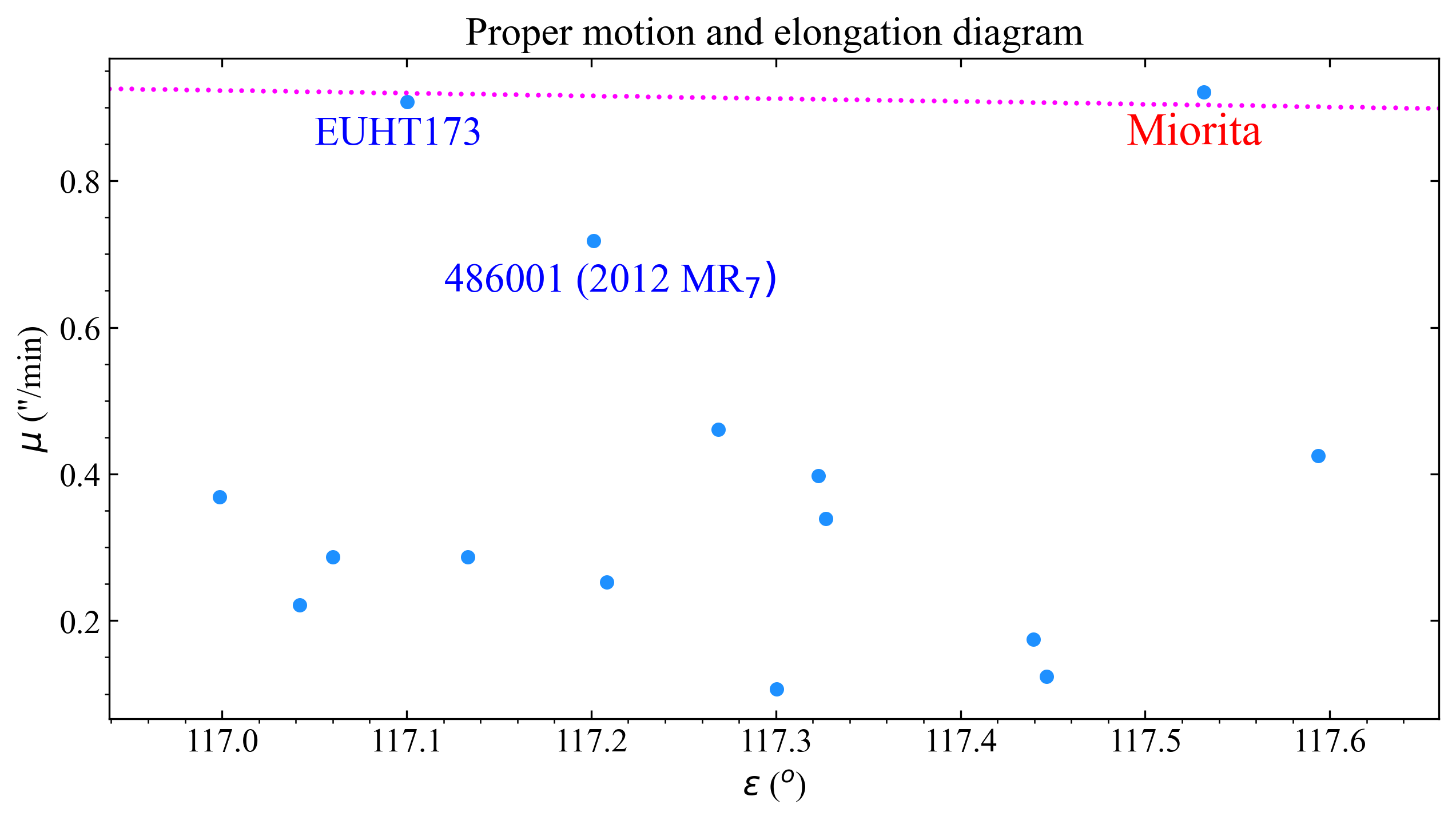}
         \includegraphics[width=\linewidth]{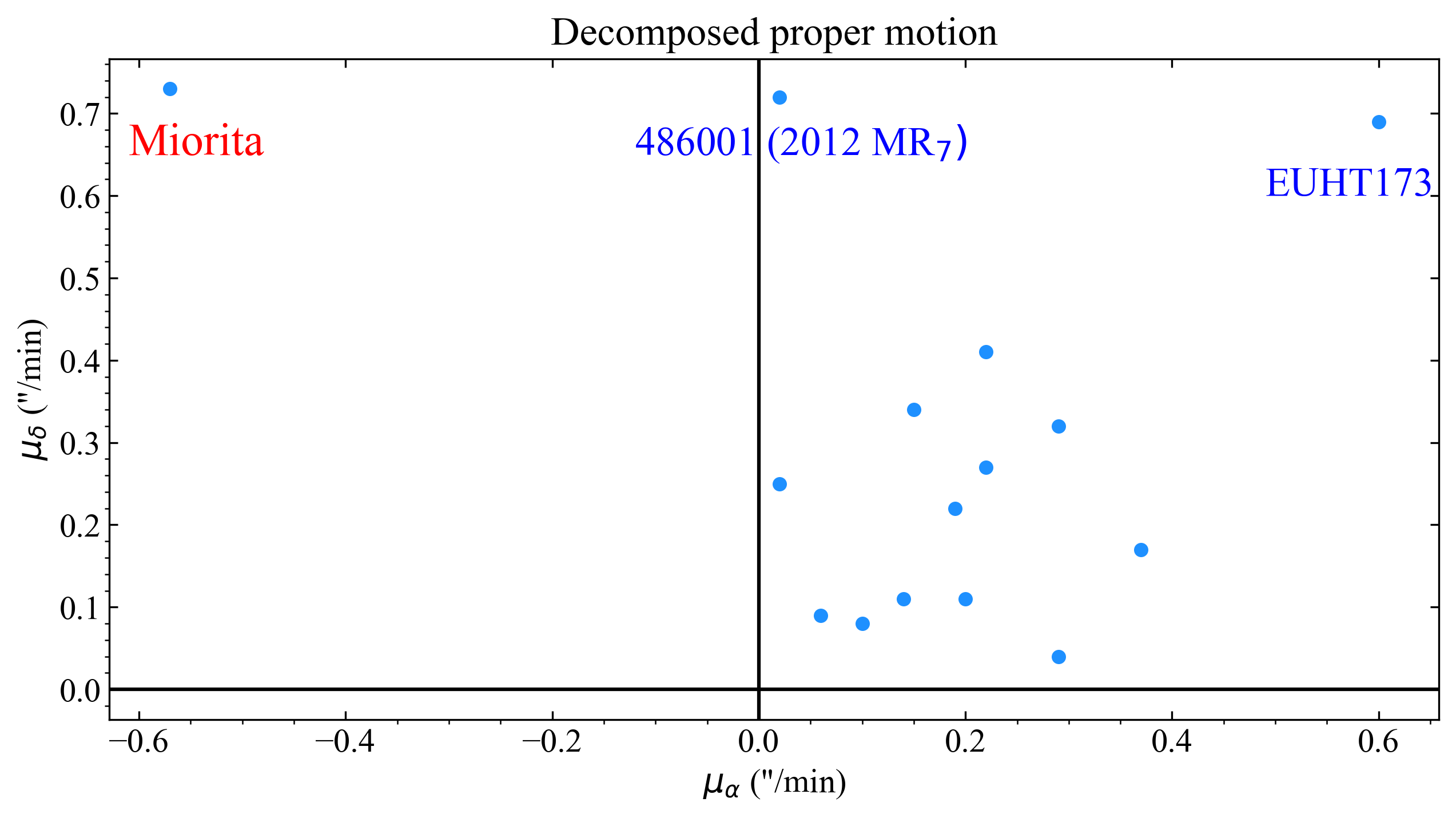}
         \includegraphics[width=\linewidth]{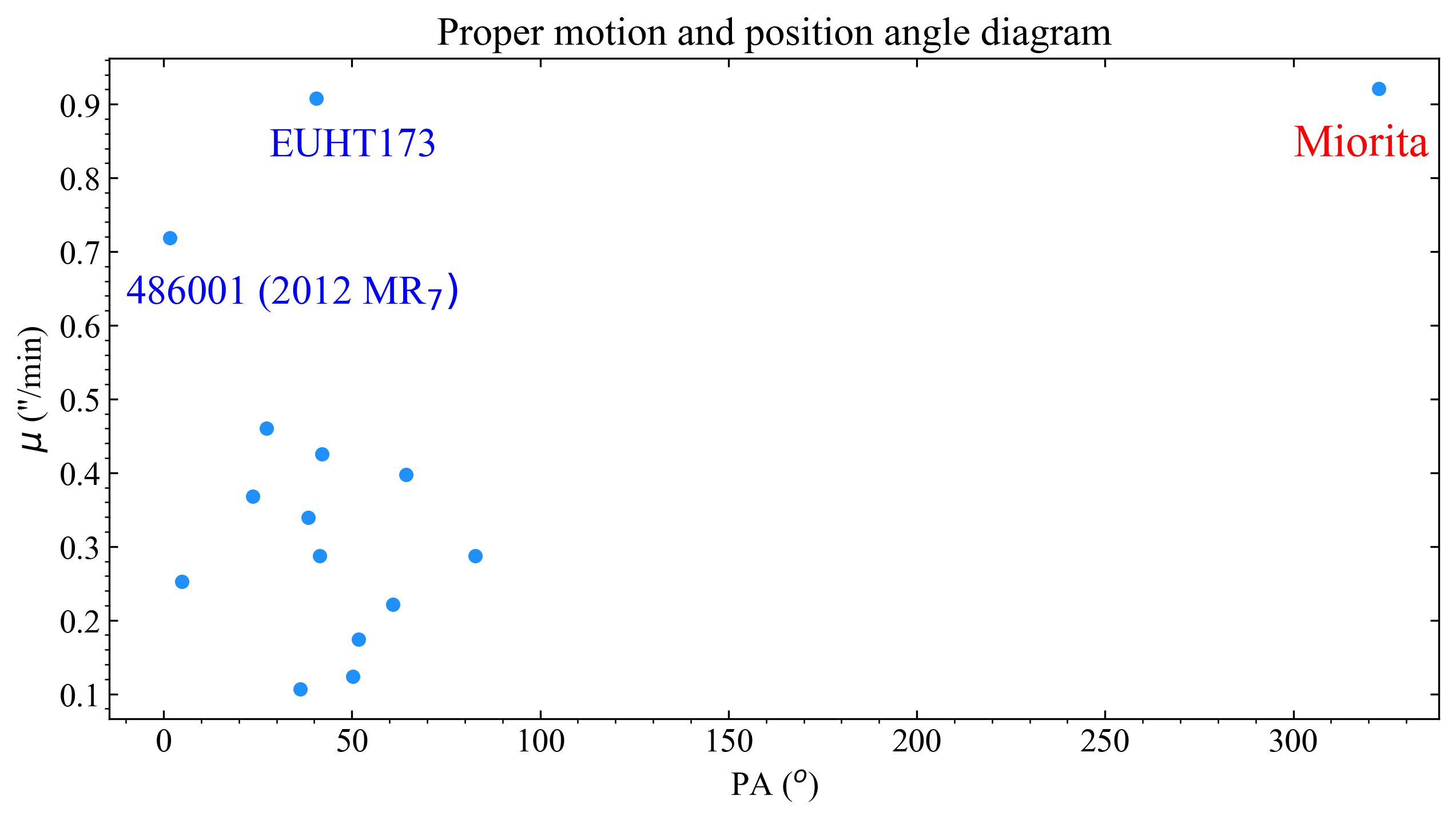}
         \caption{Observational details of the discovery of 622577 Miorita (2014~LU$_{14}$). {\it Top panel:} Results from the solar elongation--proper 
                  motion or $\epsilon$--$\mu$ tool (in magenta). {\it Middle panel:} Proper motions in equatorial coordinates. {\it Bottom panel:} 
                  Proper motions as a function of the position angle.
                 }
         \label{Miorita-run}
      \end{figure}

            As an additional quality control check, \cite{2011P&SS...59.1632V} published a simple solar elongation--proper motion or $\epsilon$--$\mu$ 
            model assuming circular and coplanar orbits, which was later implemented online in our EURONEAR {\tt NEA Checker} 
            tool\footnote{\href{http://www.euronear.org/tools/NEACheck.php}{http://www.euronear.org/tools/NEACheck.php}} aimed at assessing possible NEA candidates in any observed field. 
            Figure~\ref{Miorita-run} presents the output for this field, from which Miorita and another unknown detection clearly stand up in 
            comparison with the rest of the objects.

             With a proper motion $\mu=0.92^{\prime\prime}$/min, in the top panel of Fig.~\ref{Miorita-run}, Miorita stands just above the border line 
             (dotted magenta) which marks the separation between MBAs and NEA candidates in the $\epsilon$--$\mu$ model. The main target (known NEA 
             486001) stands below the magenta line, and another similar detection (EUHT173) stands just below the border, being identified much later 
             by the MPC as MBA 446534 (2014~MP$_{14}$) that was discovered in 2010 by the WISE survey (one month arc) and actually recovered by our 
             program (MPS~521453),

             On the middle and especially the bottom EURONEAR {\tt NEA Checker} plot in Fig.~\ref{Miorita-run}, Miorita clearly detaches as a NEA 
             candidate, particularly due to the position angle, clearly different from those of all other known and unknown MBAs (including the known 
             NEA target 486001). Actually, this direction of motion was clearly detected visually by the reducer who promptly informed the PI and 
             reported the new NEO candidate EUHT171 (which eventually became Miorita) only two and half hours after the observations were completed.

      \section{Input data and uncertainties\label{Adata}}
         Here, we include the barycentric Cartesian state vectors of 622577 Miorita (2014~LU$_{14}$) and other objects used in the calculations. These 
         vectors and their uncertainties were used to perform the calculations discussed above. As an example, a new value of the $X$-component of the 
         state vector was computed using $X_{\rm c} = X + \sigma_X \ r$, where $r$ is an univariate Gaussian random number, and $X$ and $\sigma_X$ are 
         the mean value and its 1$\sigma$ uncertainty in, e.g., Table~\ref{vectorMiorita}.
     \begin{table}
      \centering
      \fontsize{8}{12pt}\selectfont
      \tabcolsep 0.15truecm
      \caption{\label{vectorMiorita}Barycentric Cartesian state vector of 622577 Miorita (2014~LU$_{14}$): Components and associated 1$\sigma$ 
               uncertainties.
              }
      \begin{tabular}{ccc}
       \hline
        Component                         &   &    value$\pm$1$\sigma$ uncertainty                                 \\
       \hline
        $X$ (au)                          & = &    1.261525087989501$\times10^{0}$$\pm$7.81332354$\times10^{-7}$   \\
        $Y$ (au)                          & = &    1.415510074085832$\times10^{-1}$$\pm$1.65373719$\times10^{-6}$  \\
        $Z$ (au)                          & = &    6.025009389672883$\times10^{-1}$$\pm$8.41469560$\times10^{-7}$  \\
        $V_X$ (au/d)                      & = & $-$6.778246475904528$\times10^{-3}$$\pm$1.24123060$\times10^{-8}$  \\
        $V_Y$ (au/d)                      & = &    1.165360296974367$\times10^{-2}$$\pm$3.75642625$\times10^{-9}$  \\
        $V_Z$ (au/d)                      & = & $-$2.808860650274434$\times10^{-3}$$\pm$6.24561137$\times10^{-9}$  \\
       \hline
      \end{tabular}
      \tablefoot{Data are referred to epoch JD 2460800.5, which corresponds to 0:00 on 2025 May 5 TDB (J2000.0 ecliptic and equinox). Source: JPL's 
                 {\tt Horizons}.
                }
     \end{table}

     \begin{table}
      \centering
      \fontsize{8}{12pt}\selectfont
      \tabcolsep 0.15truecm
      \caption{\label{vector504181}Barycentric Cartesian state vector of 504181 (2006~TC): Components and associated 1$\sigma$ 
               uncertainties.
              }
      \begin{tabular}{ccc}
       \hline
        Component                         &   &    value$\pm$1$\sigma$ uncertainty                                 \\
       \hline
        $X$ (au)                          & = &    1.366245574867315$\times10^{0}$$\pm$6.59043937$\times10^{-7}$   \\
        $Y$ (au)                          & = &    1.502962644075712$\times10^{0}$$\pm$7.34785032$\times10^{-7}$   \\
        $Z$ (au)                          & = & $-$7.019420919109259$\times10^{-1}$$\pm$7.14488621$\times10^{-7}$  \\
        $V_X$ (au/d)                      & = & $-$8.212543579595001$\times10^{-3}$$\pm$4.84273509$\times10^{-9}$  \\
        $V_Y$ (au/d)                      & = & $-$3.030690816816023$\times10^{-3}$$\pm$5.33875719$\times10^{-9}$  \\
        $V_Z$ (au/d)                      & = &    2.328011259580962$\times10^{-3}$$\pm$2.92860933$\times10^{-9}$  \\
       \hline
      \end{tabular}
      \tablefoot{Data are referred to epoch JD 2460800.5, which corresponds to 0:00 on 2025 May 5 TDB (J2000.0 ecliptic and equinox). Source: JPL's 
                 {\tt Horizons}.
                }
     \end{table}

     \begin{table}
      \centering
      \fontsize{8}{12pt}\selectfont
      \tabcolsep 0.15truecm
      \caption{\label{vector482798}Barycentric Cartesian state vector of 482798 (2013~QK$_{48}$): Components and associated 1$\sigma$ 
               uncertainties.
              }
      \begin{tabular}{ccc}
       \hline
        Component                         &   &    value$\pm$1$\sigma$ uncertainty                                 \\
       \hline
        $X$ (au)                          & = &    1.391336477097777$\times10^{-1}$$\pm$2.32350669$\times10^{-6}$   \\
        $Y$ (au)                          & = & $-$5.133164792230683$\times10^{-1}$$\pm$1.34983145$\times10^{-6}$  \\
        $Z$ (au)                          & = &    1.057528256547908$\times10^{-1}$$\pm$2.10296029$\times10^{-7}$  \\
        $V_X$ (au/d)                      & = &    2.656004142457892$\times10^{-2}$$\pm$2.78143263$\times10^{-8}$  \\
        $V_Y$ (au/d)                      & = & $-$1.428199888293871$\times10^{-2}$$\pm$8.35894569$\times10^{-8}$  \\
        $V_Z$ (au/d)                      & = & $-$1.858328267787350$\times10^{-3}$$\pm$1.77818390$\times10^{-8}$  \\
       \hline
      \end{tabular}
      \tablefoot{Data are referred to epoch JD 2460800.5, which corresponds to 0:00 on 2025 May 5 TDB (J2000.0 ecliptic and equinox). Source: JPL's 
                 {\tt Horizons}.
                }
     \end{table}

      \section{Miorita-like orbits\label{flock}}
\begin{figure}
        \centering
         \includegraphics[width=\linewidth]{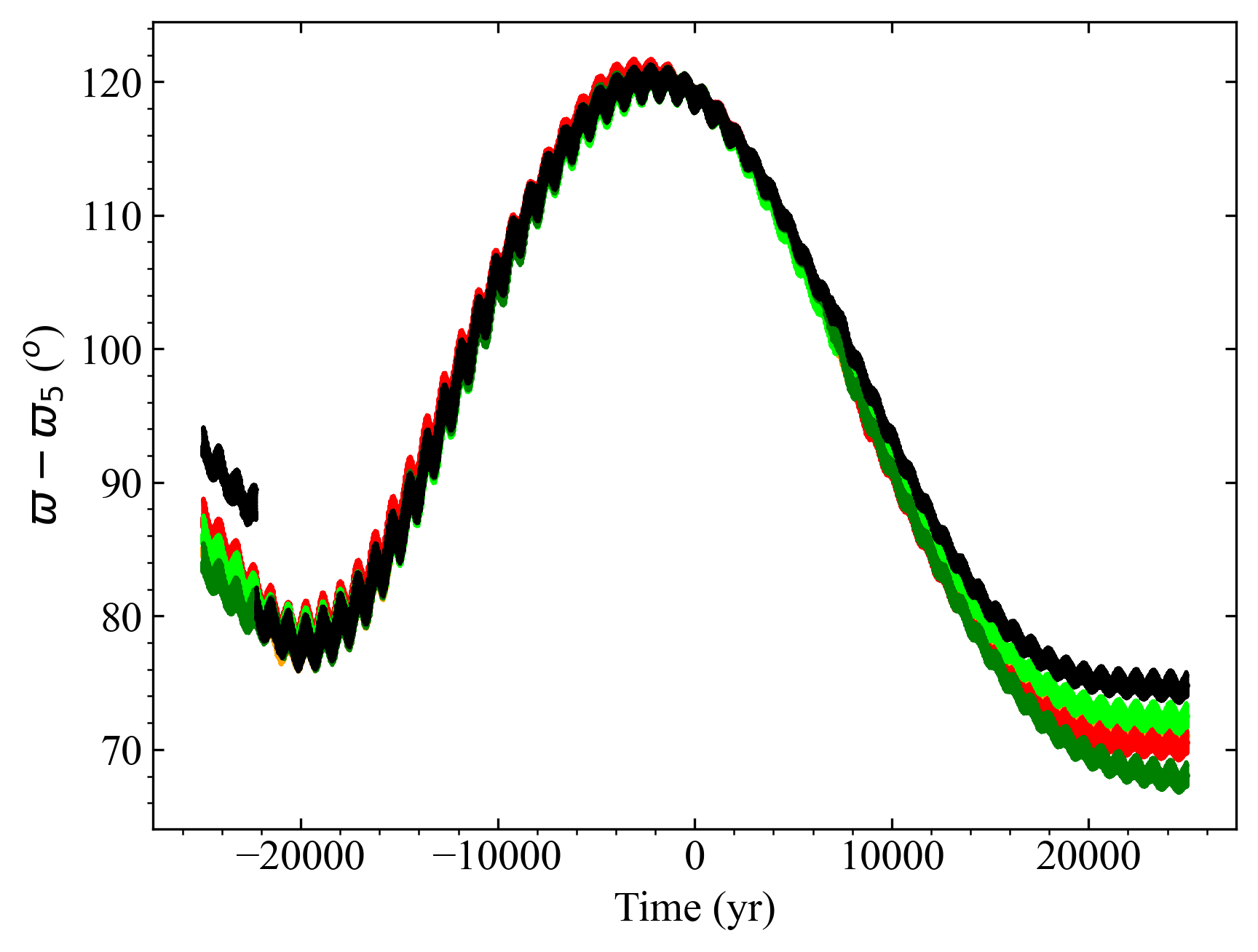}
         \caption{Time evolution of the apsidal longitude $\varpi - \varpi_{\rm 5}$ of 622577 Miorita (2014~LU$_{14}$). Results from the calculations 
                  displayed in the central panels of Fig.~\ref{Miorita}. Nominal orbit in black, those of control orbits with state vectors 
                  separated $\pm3\sigma$ from the nominal ones in lime/green, and $\pm9\sigma$ in orange/red. The source of the input data is JPL's 
                  {\tt Horizons}.
                 }
         \label{Miorita-apsidalnu5}
      \end{figure}

\begin{figure}
        \centering
         \includegraphics[width=\linewidth]{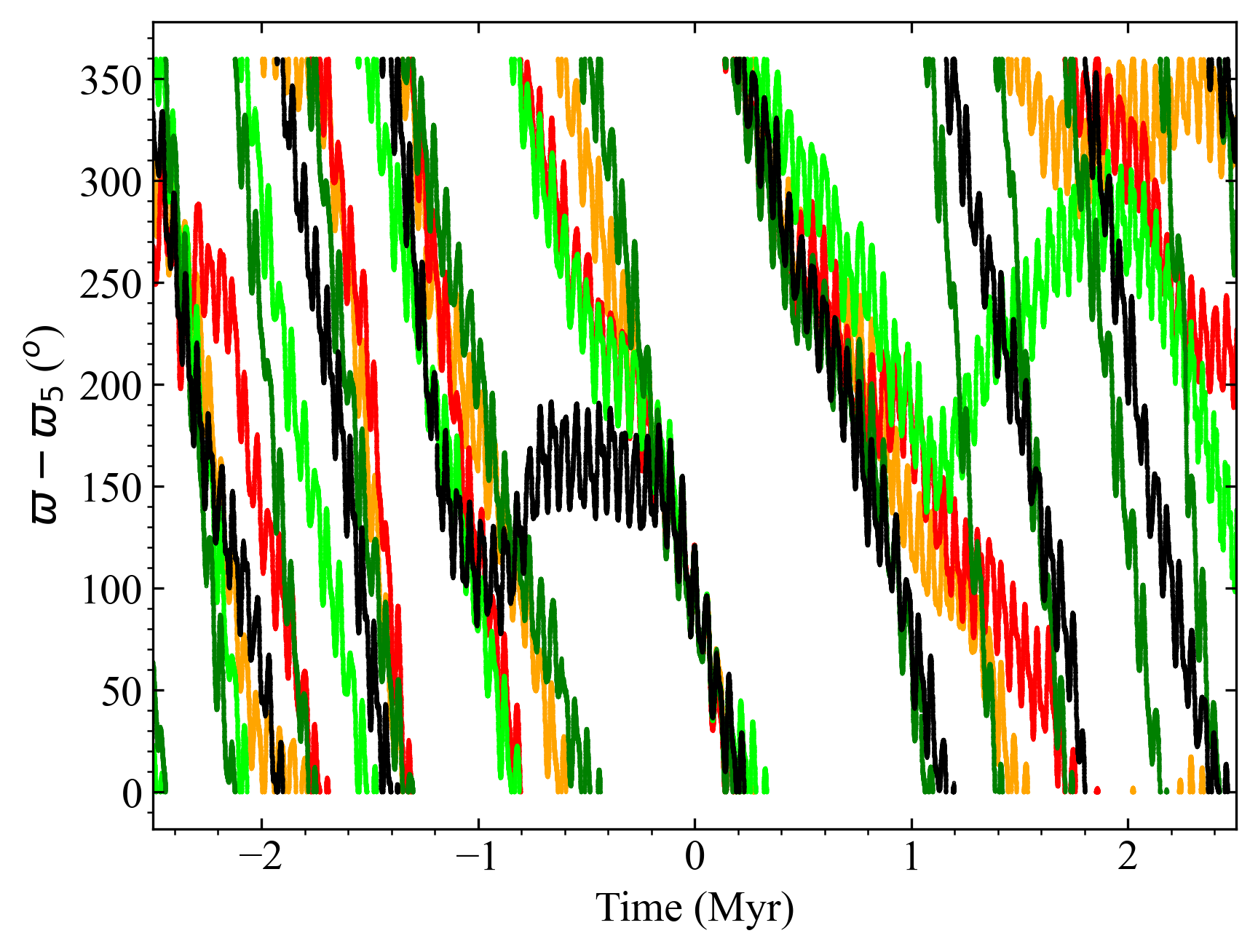}
         \caption{Time evolution of the apsidal longitude $\varpi - \varpi_{\rm 5}$ of 622577 Miorita (2014~LU$_{14}$). Results from the calculations 
                  displayed in the right panels of Fig.~\ref{Miorita}. Nominal orbit in black, those of control orbits with state vectors separated 
                  $\pm3\sigma$ from the nominal ones in lime/green, and $\pm9\sigma$ in orange/red. The source of the input data is JPL's 
                  {\tt Horizons}.
                 }
         \label{Miorita-apsidalnu5n}
      \end{figure}

         Figures~\ref{Miorita-apsidalnu5} and \ref{Miorita-apsidalnu5n} show the evolution over time of $\varpi - \varpi_{\rm 5}$ for 622577 Miorita 
         (2014~LU$_{14}$) from the same simulations presented in the central and right panels of Fig.~\ref{Miorita}, respectively. The short-term 
         evolution in Fig.~\ref{Miorita-apsidalnu5} shows an oscillation of $\varpi - \varpi_{\rm 5}$ that hints at an apsidal resonance with Jupiter 
         for the control orbits, but the longer-term evolution in Fig.~\ref{Miorita-apsidalnu5n} is more consistent with a near apsidal resonance with 
         episodes of actual capture.

\begin{figure*}
        \centering
         \includegraphics[width=0.24\linewidth]{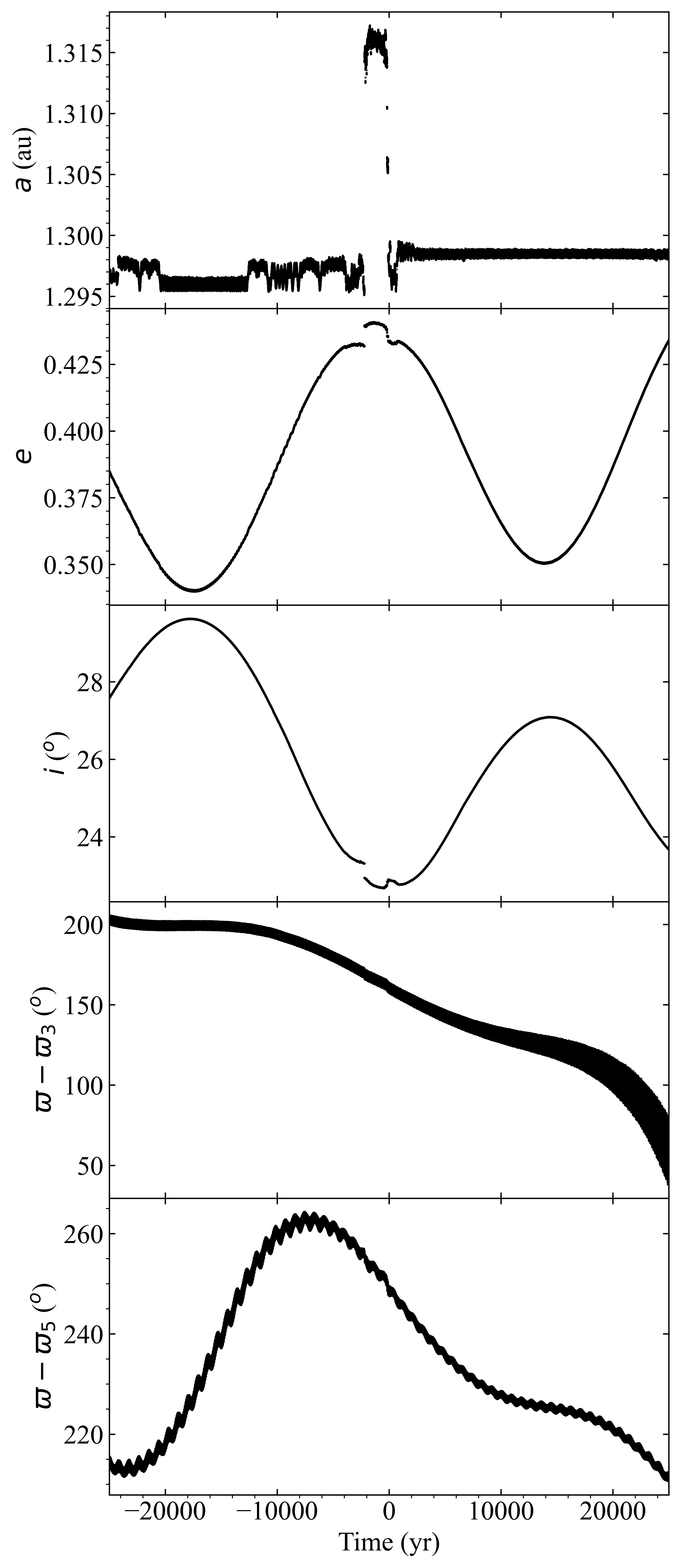}
         \includegraphics[width=0.24\linewidth]{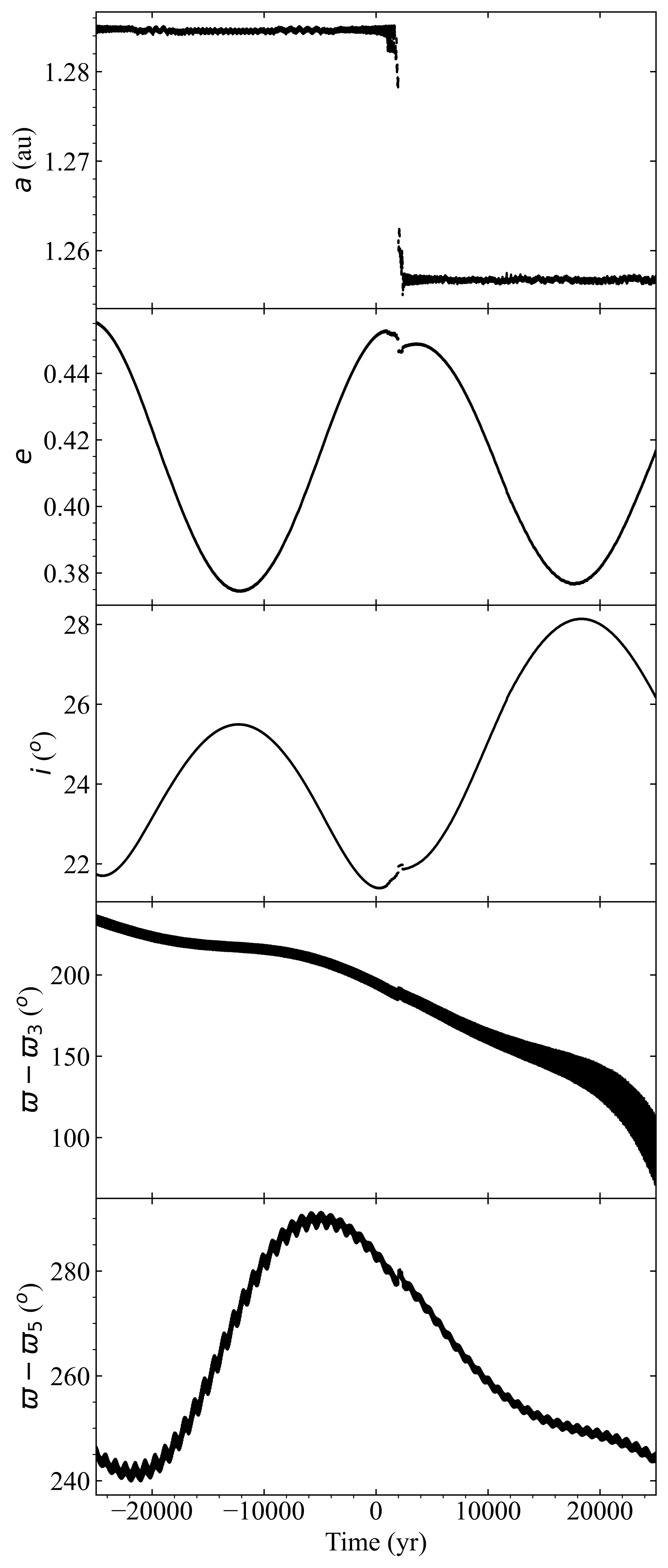}
         \includegraphics[width=0.24\linewidth]{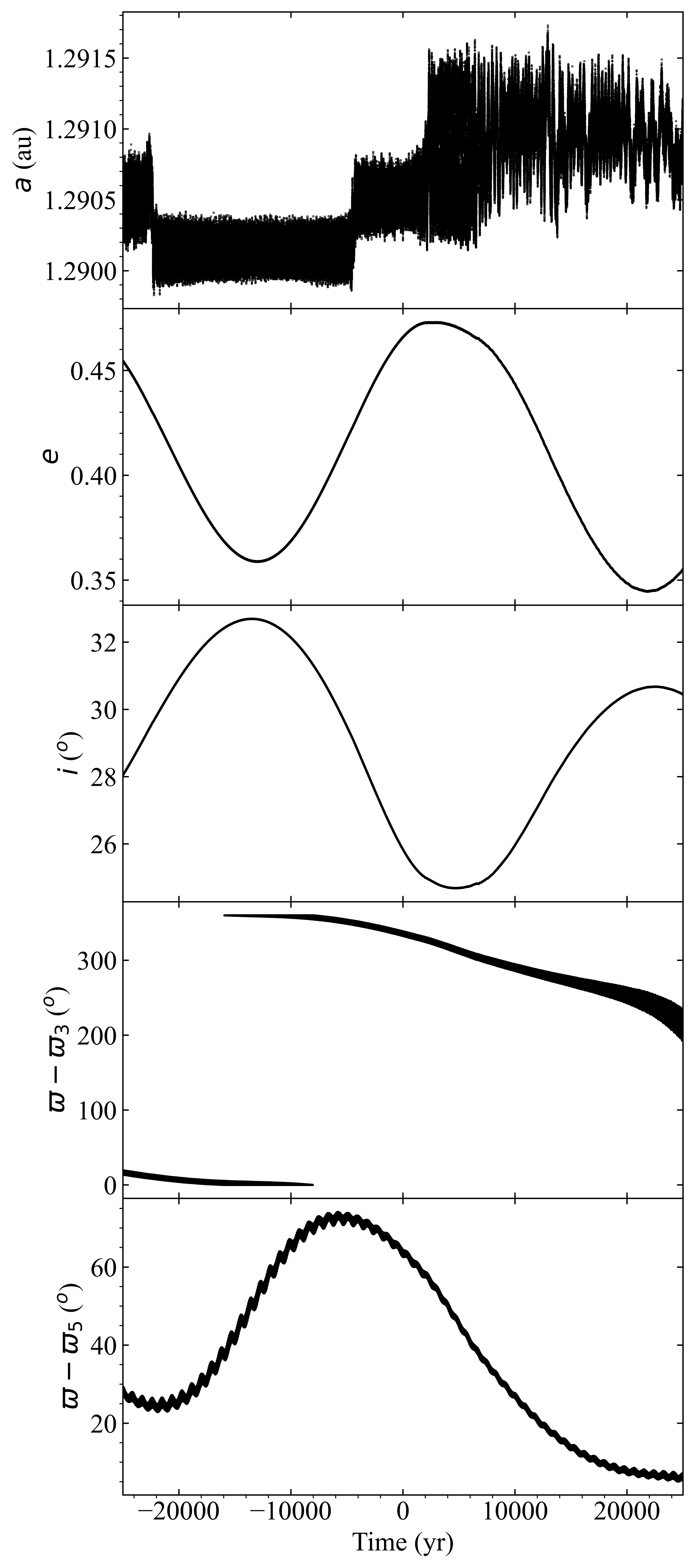}
         \includegraphics[width=0.24\linewidth]{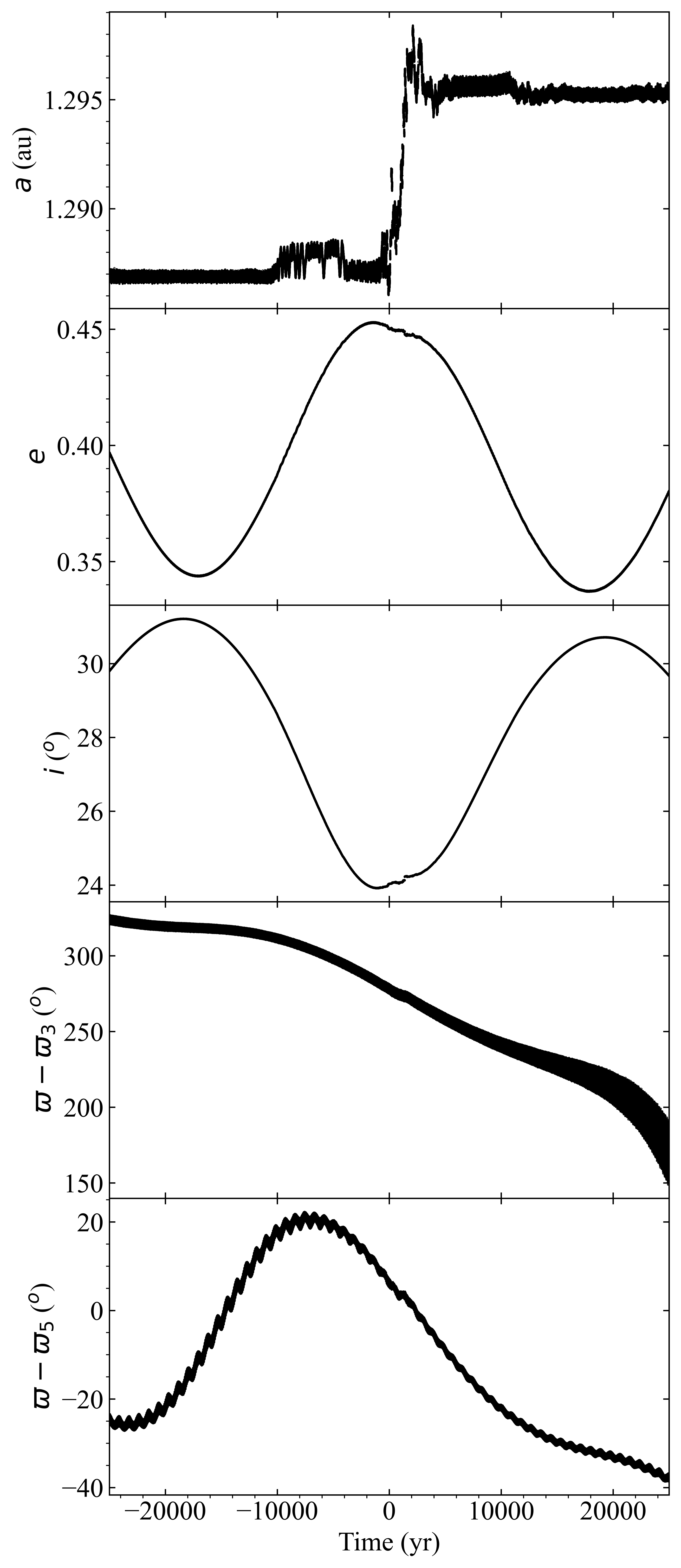}
         \caption{Orbital evolution of NEAs 387668 (2002~SZ), 2004~US$_{1}$, 299582 (2006~GQ$_{2}$), and 2018~AC$_{4}$. {\it Left panels:} Past and 
                  future evolution of the nominal orbit of 387668. {\it Second to left panels:} Past and future evolution of the nominal orbit of 
                  2004~US$_{1}$. {\it Second to right panels:} Past and future evolution of the nominal orbit of 299582. {\it Right panels:} Past and 
                  future evolution of the nominal orbit of 2018~AC$_{4}$. Time evolution of the value of the semimajor axis, $a$ (top panels), 
                  eccentricity, $e$ (second to top), inclination, $i$ (middle), the apsidal longitude relative to Earth, $\varpi - \varpi_{\rm 3}$
                  (second to bottom), and the apsidal longitude relative to Jupiter, $\varpi - \varpi_{\rm 5}$ (bottom). The origin of time is the 
                  epoch 2460800.5~JD Barycentric Dynamical Time (2025-May-05.0 00:00:00.0 TDB) and the output cadence is 36.525~d.  
                  The source of the input data is JPL's {\tt Horizons}. 
                 }
         \label{Miorita-flock}
      \end{figure*}

         Figure~\ref{Miorita-flock} shows the evolution of NEAs 387668 (2002~SZ), 2004~US$_{1}$, 299582 (2006~GQ$_{2}$), and 2018~AC$_{4}$, whose 
         current orbits resemble that of Miorita. As Miorita, they are subjected to a von Zeipel-Lidov-Kozai secular resonance, but they are also in a 
         near apsidal resonance, both controlled by Jupiter.

     \section{Long-term evolution of Miorita-like orbits\label{beyond}}
\begin{figure*}
        \centering
         \includegraphics[width=0.32\linewidth]{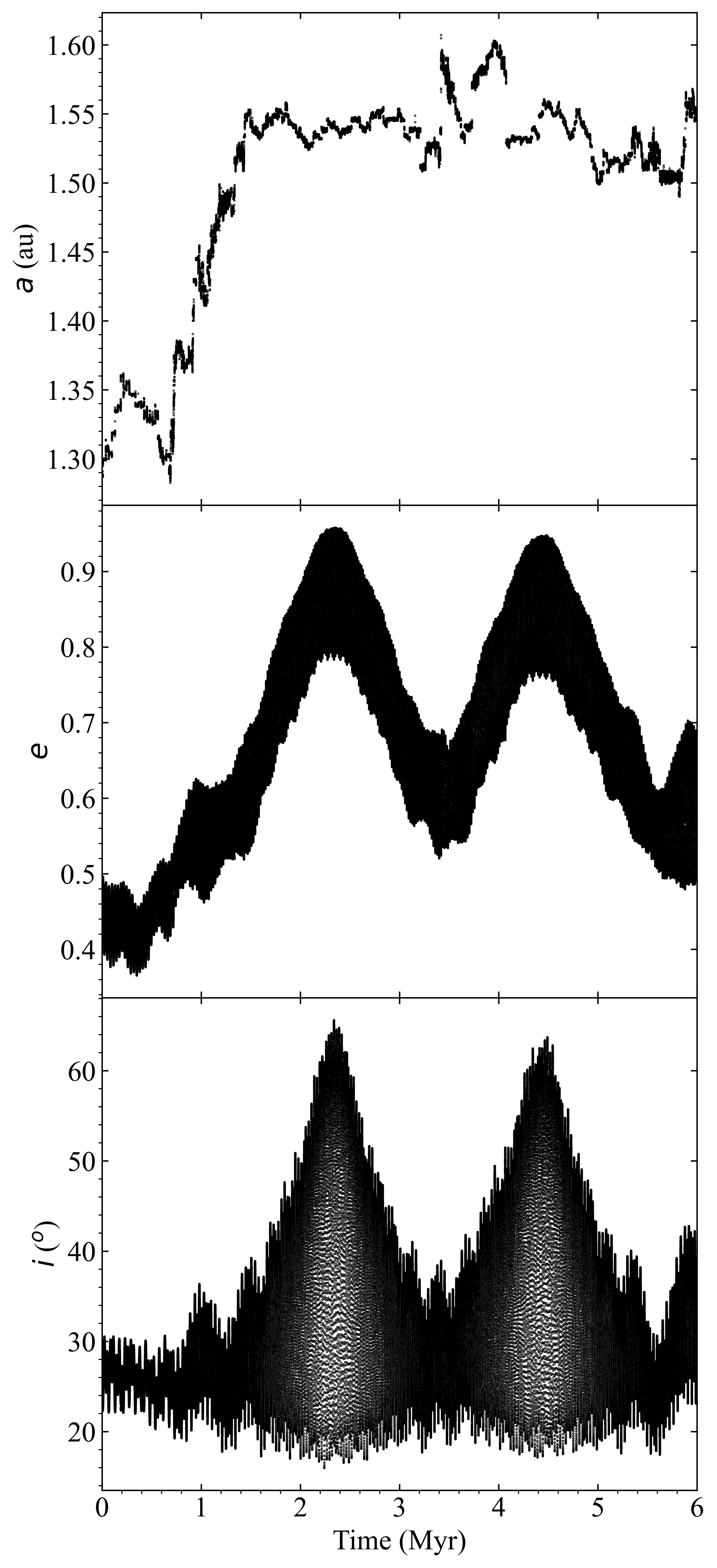}
         \includegraphics[width=0.32\linewidth]{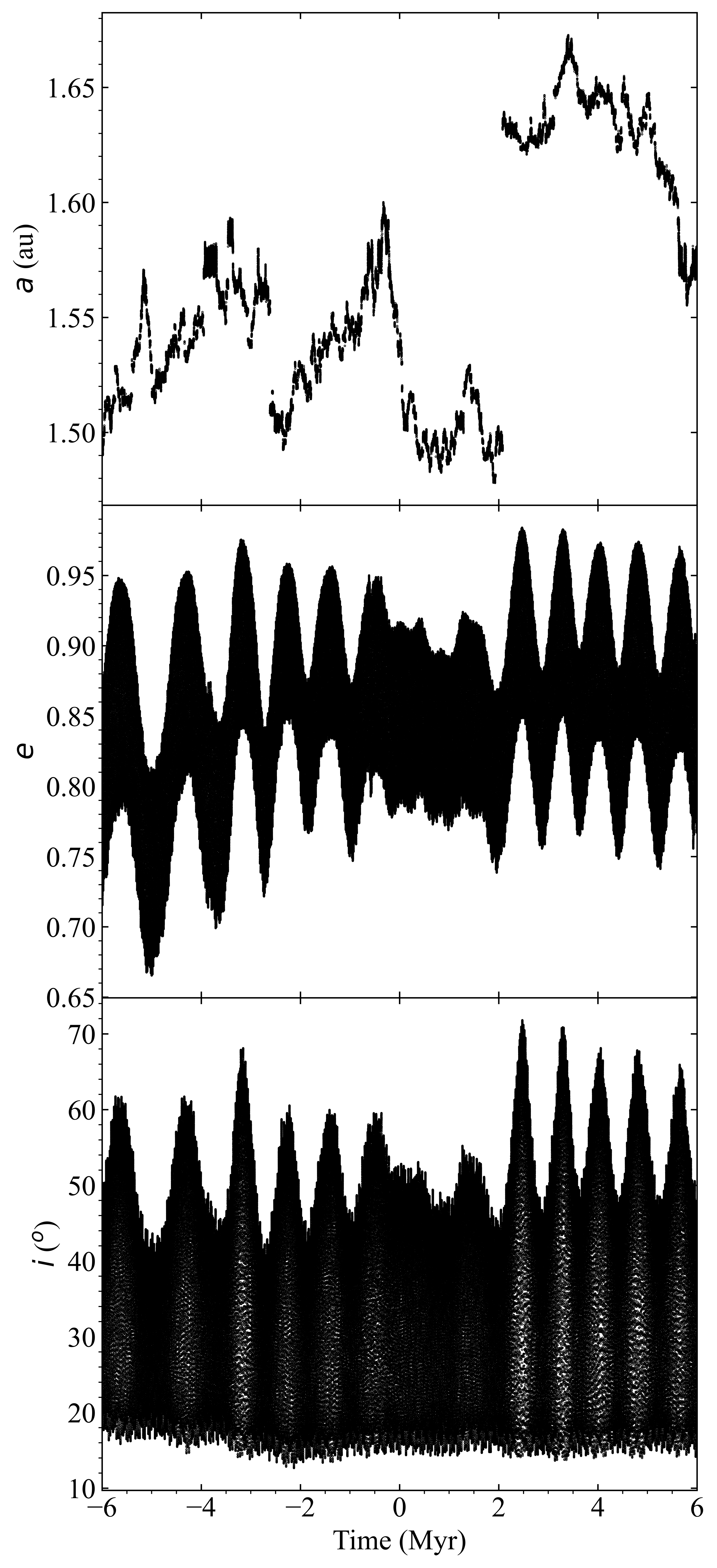}
         \includegraphics[width=0.32\linewidth]{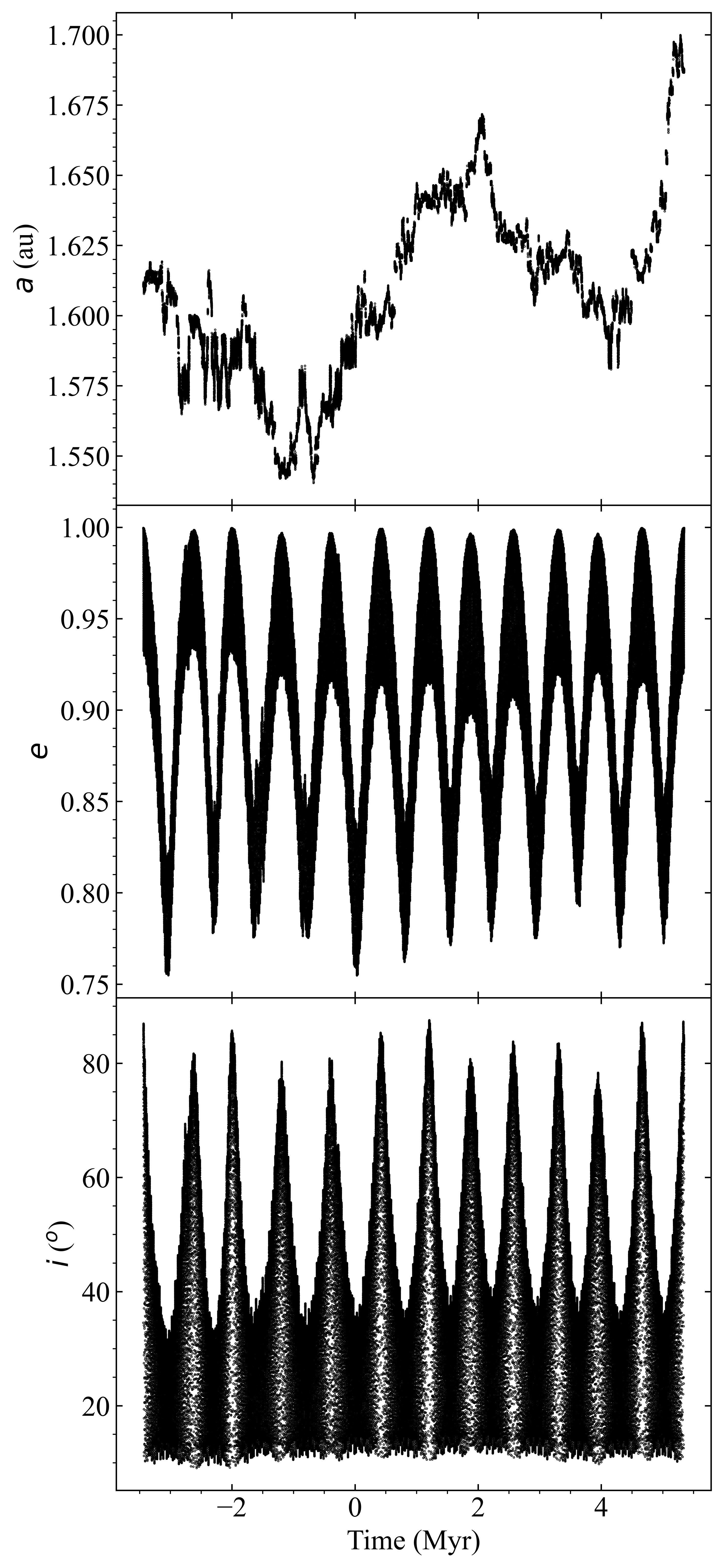}
         \caption{Orbital evolution of a representative control orbit of 622577 Miorita (2014~LU$_{14}$) and NEAs 504181 (2006~TC), and 482798
                  (2013~QK$_{48}$). {\it Left panels:} Future evolution of the control orbit with state vectors separated $-3\sigma$ from the nominal 
                  ones, this relevant clone appears in lime in the right panels of Fig.~\ref{Miorita}. {\it Central panels:} Past and future evolution 
                  of the nominal orbit of 504181. {\it Right panels:} Past and future evolution of the nominal orbit of 482798. Time evolution of the 
                  value of the semimajor axis, $a$ (top panels), eccentricity, $e$ (middle), inclination, $i$ (bottom). The origin of time is the 
                  epoch 2460800.5~JD Barycentric Dynamical Time (2025-May-05.0 00:00:00.0 TDB) and the output cadence is 36.525~d for the left and 
                  central panels, and 25~yr for the right panels. The source of the input data is JPL's {\tt Horizons}.
                 }
         \label{apsidal}
      \end{figure*}

        Figure~\ref{apsidal}, left panels, shows the evolution into the future of the control orbit with state vectors separated $-3\sigma$ from the 
        nominal ones, this relevant clone appears in lime in the right panels of Fig.~\ref{Miorita}. It shows correlated long-period ($\sim$2~Myr) 
        oscillations in the values of $e$ and $i$, middle and bottom panels, respectively. These oscillations are linked to the apsidal resonance with 
        Jupiter discussed in Sect.~\ref{Discussion}. Figure~\ref{apsidal} also shows the past and future evolution of the nominal orbits (see 
        Table~\ref{elementso}) of 504181 (2006~TC), central panels, and 482798 (2013~QK$_{48}$), right panels, the input state vectors appear in
        Tables~\ref{vector504181} and \ref{vector482798}. \citet{2024PSJ.....5..113M} have suggested that 504181 might be the largest fragment of the 
        asteroid breakup that created the Phaethon--Geminid complex. Apollo-class NEA 482798 is a PHA that experiences relatively close encounters 
        with Mars, Venus and Mercury. In Fig.~\ref{apsidal}, the nominal orbit of 482798 goes into the Sun both integrating forward and backwards.  
      \begin{table*}
         \centering
         \fontsize{8}{11pt}\selectfont
         \tabcolsep 0.15truecm
         \caption{\label{elementso}Heliocentric Keplerian orbital elements of asteroids 504181 (2006~TC) and 482798 (2013~QK$_{48}$). 
                 }
         \begin{tabular}{lccc}
            \hline\hline
             Parameter                                         &   &    504181                    &   482798                    \\
            \hline
             Semimajor axis, $a$ (au)                          & = &    1.53827035$\pm$0.00000002 &   1.58573770$\pm$0.00000002 \\
             Eccentricity, $e$                                 & = &    0.91204242$\pm$0.00000011 &   0.82973801$\pm$0.00000007 \\
             Inclination, $i$ (\degr)                          & = &   19.54779$\pm$0.00002       &   18.967424$\pm$0.000012    \\
             Longitude of the ascending node, $\Omega$ (\degr) & = &  151.76208$\pm$0.00011       & 141.39855$\pm$0.00002       \\
             Argument of perihelion, $\omega$ (\degr)          & = &   61.51161$\pm$0.00010       & 47.18181$\pm$0.00002        \\
             Mean anomaly, $M$ (\degr)                         & = &  290.80622$\pm$0.00006       & 8.47648$\pm$0.00005         \\
             Perihelion, $q$ (au)                              & = &    0.1353025$\pm$0.0000002   &   0.26999085$\pm$0.00000011 \\
             Aphelion, $Q$ (au)                                & = &    2.94123816$\pm$0.00000004 &   2.90148454$\pm$0.00000004 \\
             MOID with the Earth (au)                          & = &    0.144222                  &   0.026112                  \\
             Absolute magnitude, $H$ (mag)                     & = &   18.81                      &  18.35                      \\ 
             Data-arc span (d)                                 & = & 4792                         &   8520                      \\
             Number of observations                            & = &   59                         & 149                         \\
            \hline
         \end{tabular}
         \tablefoot{Values include the 1$\sigma$ uncertainty. The orbits of 504181 (solution date, March 1, 2023, 06:49:57 PST) and 482798 (solution 
                    date, December 22, 2024, 05:28:42 PST) are referred to epoch JD 2460800.5, which corresponds to 0:00 on 2025 May 5 TDB 
                    (Barycentric Dynamical Time, J2000.0 ecliptic and equinox). Source: JPL's SBDB.
                   }
      \end{table*}

   \end{appendix}

\end{document}